\newcommand{\lya}{\mbox{Ly$\alpha$}}

\newcommand{\halpha}{\mbox{H$\alpha$}}
\newcommand{\hbeta}{\mbox{H$\beta$}}
\newcommand{\hgamma}{\mbox{H$\gamma$}}
\newcommand{\hdelta}{\mbox{H$\delta$}}
\newcommand{\hepsilon}{\mbox{H$\epsilon$}}

\newcommand{\hII}{\mbox{H~{\sc ii}}}
\newcommand{\heI}{\mbox{He~{\sc i}}}
\newcommand{\heII}{\mbox{He~{\sc ii}}}
\newcommand{\oI}{\mbox{O~{\sc i}}}
\newcommand{\oII}{\mbox{O~{\sc ii}}}
\newcommand{\oIII}{\mbox{O~{\sc iii}}}

\newcommand{\cII}{\mbox{C~{\sc ii}}}
\newcommand{\cIII}{\mbox{C~{\sc iii}}}
\newcommand{\cIV}{\mbox{C~{\sc iv}}}
\newcommand{\neIII}{\mbox{Ne~{\sc iii}}}

\newcommand{\sII}{\mbox{S~{\sc ii}}}

\newcommand{\nII}{\mbox{N~{\sc ii}}}
\newcommand{\nIII}{\mbox{N~{\sc iii}}}
\newcommand{\nIV}{\mbox{N~{\sc iv}}}

\newcommand{\mgII}{\mbox{Mg~{\sc ii}}}

\newcommand{\op}{\mbox{O$^{+}$}}
\newcommand{\opp}{\mbox{O$^{++}$}}

\newcommand{\kms}{\mbox{km\,s$^{-1}$}}

\newcommand{\msun}{\mbox{M$_\odot$}}

\newcommand{\fesclyc}{\mbox{$f_\mathrm{esc}^\mathrm{LyC}$}}

\newcommand{\fescmgii}{\mbox{$f_\mathrm{esc}^{\mathrm{MgII}}$}}

\newcommand{\etemp}{\mbox{$T_\mathrm{e}$}}
\newcommand{\edens}{\mbox{$n_\mathrm{e}$}}

\newcommand{\Muv}{\mbox{$M_\mathrm{UV}$}}

\newcommand{\ewoIII}{\mbox{$W_{\mathrm{OIII}}$}}
\newcommand{\ewcIII}{\mbox{$W_{\mathrm{CIII]}}$}}
\newcommand{\ewcIV}{\mbox{$W_{\mathrm{CIV}}$}}

\newcommand{\xiion}{\mbox{$\xi_{\mathrm{ion}}$}}

\newcommand{\logOH}{\mbox{$12+\log(\mathrm{O/H})$}}
\newcommand{\logCO}{\mbox{$\log(\mathrm{C/O})$}}
\newcommand{\logNO}{\mbox{$\log(\mathrm{N/O})$}}

\newcommand{\logU}{\mbox{$\log(U)$}}

\newcommand{\nodata}{...}

\documentclass[a4paper,fleqn,usenatbib,twocolumn]{aastex631}
\usepackage{newtxtext,newtxmath}
\usepackage[T1]{fontenc}
\usepackage{xcolor}
\usepackage{longtable}
\usepackage{tablefootnote}
\usepackage{graphicx}   
\usepackage{amsmath}    

\received{Int' nu}
\revised{iiinntt' nuuu..}
\accepted{Men Nu!}
\submitjournal{ApJ}

\shorttitle{Stack the DJA}
\shortauthors{M. J. Hayes et al.}

\begin{document}

\title{ON THE AVERAGE ULTRAVIOLET EMISSION LINE SPECTRA OF HIGH-REDSHIFT GALAXIES: \\
HOT AND COLD, CARBON-POOR, NITROGEN MODEST, AND OOZING IONIZING PHOTONS}

\correspondingauthor{Matthew J. Hayes}
\email{matthew.hayes@astro.su.se}

\author[0000-0001-8587-218X]{Matthew J. Hayes}
\affiliation{Stockholm University, Department of Astronomy and Oskar Klein Centre for Cosmoparticle Physics, AlbaNova University Centre, SE-10691, Stockholm, Sweden.}

\author[0000-0001-8419-3062]{Alberto Saldana-Lopez}
\affiliation{Stockholm University, Department of Astronomy and Oskar Klein Centre for Cosmoparticle Physics, AlbaNova University Centre, SE-10691, Stockholm, Sweden.}

\author[0009-0000-9676-0538]{Annalisa Citro}
\affiliation{Minnesota Institute for Astrophysics, School of Physics and Astronomy, University of Minnesota, 316 Church Str. SE, Minneapolis, MN 55455, USA.}

\author[0000-0003-4372-2006]{Bethan L. James}
\affiliation{AURA for ESA, Space Telescope Science Institute, 3700 San Martin Drive, Baltimore, MD 21218, USA.}

\author[0000-0003-2589-762X]{Matilde Mingozzi}
\affiliation{AURA for ESA, Space Telescope Science Institute, 3700 San Martin Drive, Baltimore, MD 21218, USA.}

\author[0000-0002-9136-8876]{Claudia Scarlata}
\affiliation{Minnesota Institute for Astrophysics, School of Physics and Astronomy, University of Minnesota, 316 Church Str. SE, Minneapolis, MN 55455, USA.}

\author[0009-0000-2997-7630]{Zorayda Martinez}
\affiliation{Department of Astronomy, The University of Texas at Austin, 2515 Speedway, Stop C1400, Austin, TX 78712, USA.}

\author[0000-0002-4153-053X]{Danielle A. Berg}
\affiliation{Department of Astronomy, The University of Texas at Austin, 2515 Speedway, Stop C1400, Austin, TX 78712, USA.}

\begin{abstract}
We determine the spectroscopic properties of $\simeq 1000$ ostensibly star-forming galaxies at redshifts ($z=4-10$) using prism spectroscopy from JWST/NIRSpec. With rest-wavelength coverage between \lya\ and [\sII] in the optical, we stack spectra as a function of nebular conditions, and compare UV spectral properties with stellar age.  This reveals UV lines of \nIII], \nIV], \cIII], \cIV, \heII, and \oIII] in the average high-$z$ galaxy.  All the UV lines are more intense in younger starbursts.  We measure electron temperatures from the collisionally excited [\oIII] line ratios, finding $T_\mathrm{e}=18000-22000$~K for the \opp\ regions.  We also detect a significant nebular Balmer Jump, from which we estimate only $T_\mathrm{e}=8000-13000$~K. Accounting for typical temperature offsets between zones bearing doubly and singly ionized oxygen, these two temperatures remain discrepant by around 40\%.  We use the [\oIII] temperatures to estimate abundances of carbon, nitrogen, and oxygen.  We find that log(C/O) is consistently $\simeq-1$, with no evolution of C/O with metallicity or stellar age.  The average spectra are mildly enhanced in Nitrogen, with higher N/O than low-$z$ starbursts, but are less enhanced than samples of recently reported, high-$z$, extreme galaxies that show \nIII] and \nIV] emission in the UV.  Whatever processes produce the N-enhancement in the individual galaxies must also be ongoing, at lower levels, in the median galaxy in the early Universe.  The strongest starbursts are a source of significant ionizing emission: ionizing photon production efficiencies reach $10^{25.7}$~Hz\,erg$^{-1}$, and show multiple signatures of high Lyman continuum escape, including \mgII\ escape fractions nearing 100\%, significant deficits in [\sII] emission, high degrees of ionization, and blue UV colors. 
\end{abstract}

\keywords{ Galaxies: high-redshift --- Galaxies: abundances ---  Galaxies: ISM --- Galaxies: emission lines}

\section{Introduction}\label{sect:intro}

The James Webb Space Telescope (JWST) has revolutionized our understanding of early galaxy formation.  Over the last two years, data from the telescope have updated our view on the number density of galaxies at the dawn of cosmic star-formation \citep[e.g.][]{Bouwens.2023, Donnan.2023, Finkelstein.2024, Harikane.2023lfs, Robertson.2024}, and enabled the intriguing discovery of massive-black-hole galaxy hosts \citep[e.g.][]{Maiolino.2023agn,Kocevski.2024,Kokorev.2024,Scholtz.2024}, as well as dusty, quenched systems \citep{Barrufet.2024, Carnall.2024,Weibel.2024}. It has allowed us to  witness the early chemical enrichment of galaxies \citep[e.g.][]{Arellano-Cordova.2022, Curti.2023, Schaerer.2022b.1stJWST, Trump.2023, Stiavelli.2023}, as well as the ionization history and topology of the Universe using the \lya\ emission line \citep[e.g.][]{HayesScarlata.2023, Bruton.2023.GNz11, Bruton.2023.LF, Tang.2023, Tang.2024,Witstok.2024, Napolitano.2024}. 

With the advent of NIRSpec spectroscopy \citep{Jakobsen.2022}, the first statistical samples of high-$z$ galaxies with rest-optical spectroscopy have been assembled. Emission line ratio diagrams  \citep[e.g.][]{Backhaus.2024,Cameron.2023, Mascia.2023, Shapley.2024, Sanders.2024}, suggest gas to be highly excited, due to illumination by a hard ionizing spectrum. Moreover, access to emission lines whose ratios are temperature-sensitive has enabled the determination of robust estimates of oxygen abundances, facilitating new metallicity calibrations \citep[e.g.][]{Sanders.2024, Curti.2024} and therefore the determination of the mass-metallicity relation (MZR) for the first time at $z \geq 4$ \citep[e.g.][]{He.2024,Nakajima.2023,Chemerynska.2024}.  At these low masses (or high redshifts), not only does the normalization of the MZR decrease, but its slope also flattens, which may hold substantial implications for our understanding of the baryon cycle of early galaxies. 

However, the rest-ultraviolet wavelength range provides additional and even-more-valuable constraints on the nebular conditions of the warm and hot gas \citep[e.g.][]{Berg.2019.CIVHeII, Senchyna.2022}. UV diagnostic diagrams are pivotal to break the degeneracies inherent in classical optical diagnostics \citep[e.g.][]{Feltre.2016, Jaskot.2016,Nakajima.2018,Mingozzi.2022,Mingozzi.2024}, and have the potential to reveal the true, underlying ionizing mechanisms. Some UV lines ratios provide insights on the physical conditions (density and temperature) of the more highly ionized media (where the gas is denser and highly ionized), and at densities above the critical densities of some of the strong optical diagnostics.  Finally, carbon and nitrogen emission lines can be relatively strong in the UV but weak or absent in the optical, providing information on the chemical evolution/abundance patterns of early galaxies. 

JWST has made substantial progress in constraining the chemical enrichment of high-$z$ galaxies. Carbon abundances (usually given as C/O) have been studied from the ground \citep[e.g.][]{Erb.2010,Stark.2014,Mainali.2020, Matthee.2021, Citro.2024} upon which JWST could build \citep[e.g.][]{Arellano-Cordova.2022,Curti.2023,Isobe.2023,Jones.2023}, usually finding C/O ratios consistent with the local universe at fixed (and usually low) oxygen abundance.  Perhaps even more interestingly, JWST spectra have identified a handful of `nitrogen-loud' galaxies, of which seven are currently known: GNz11 \citep{Bunker.2023},  GLASS\_150008 \citep{Isobe.2023}, CEERS-1019 \citep{Marques-Chaves.2024}, GS3073 \citep[][originally found by \citealt{Vanzella.2010}]{Ji.2024}, GHZ2 \citep{Castellano.2024}, GHZ9 \citep{Napolitano.2024_GHZ9}, RXCJ2248-4431 \citep{Topping.2024}, and GN-z9p4 \citep{Schaerer.2024}. These objects are peculiar, and seem to be absent at lower redshifts, although the noteworthy exception of Mrk\,996 \citep{James.2009} implies this is not uniquely true.  While nitrogen should follow qualitatively similar primary (core-collapse supernova) and secondary (low-intermediate mass late stages/AGB) enrichment patterns as carbon, it is difficult to explain the combined finding that carbon is lacking but nitrogen is enhanced \citep[e.g.][]{Berg.2019.CNOdwarf,Berg.2020}.   These objects have therefore triggered a lot of speculation about the reasons for their nitrogen luminosity, including exotic scenarios such as supermassive stars produced by stellar mergers in dense star clusters \citep[e.g.][]{Charbonnel.2023}, AGN activity \citep[e.g.][]{Maiolino.2024gnz11}, Tidal Disruption Events \citep{Cameron.2023} or carefully balanced recent star-formation histories \citep{Kobayashi.2024}. 

Finally, many of the objects discussed above for which detailed emission line spectroscopy has been obtained lie at redshifts where the Universe is demonstrated to be substantially neutral.  Some of the same features used for classification and abundance inferences are also key estimators of both the intrinsic ionizing power of a galaxy (e.g. the Balmer series of emission lines) and the escape of ionizing radiation (e.g. [\oIII]5008, [\oII]3727, [\sII]6717,31, \mgII2800). At these redshifts, the Universe is fully opaque to ionizing radiation \citep[LyC,][]{Inoue.2014} and the ionizing power of galaxies can never be observed directly.  Thus the mentioned emission lines are critical observables in the reionization epoch, as they provide indirect diagnostics of the amount of ionizing radiation escaping galaxies \citep[\fesclyc, E.g.][]{Leitet.2013,Izotov.2016bLyc,Flury.2022pap2,Chisholm.2022}. 

As exciting as these results are, they are mostly restricted to small samples of individual galaxies (usually one per paper, where nitrogen enhancement is concerned) that are identified by carefully inspecting archival spectra and publishing the gems.  Given the small rest-frame EWs of these UV emission lines (compared to their optical counterparts) they are harder to detect -- this causes JWST to lag somewhat in the delivery of UV-based inferences compared to more traditional methods using stronger lines in the optical. The uncertainties on these metal abundances are often large, and it is likely that there is little to be gained by publishing large numbers of limits on (C/O) and (N/O) for the plethora of non-detections.  Thus we have little insight into the distribution of these abundances across the majority of the iceberg that remains submerged.

There is one obvious way out of this: to co-add the restframe spectra of large numbers of galaxies, in a technique known as a stacking analysis.  Stacking has, for example, allowed \citet{Hu.2024} to produce a composite spectrum of 63 high-redshifts galaxies observed with NIRSpec in medium resolution gratings to study metallicities, dust attenuation, and various other properties of the ionized gas.  They conclude that average $>5.5$ galaxies are highly ionized and deficient in carbon, similarly to the remarks above.  \citet{Roberts-Borsani.2024} similarly stack $\simeq 500$ galaxies observed with the lower resolution prism and study their redshift evolution, finding them to be bluer and younger at higher redshifts and deficient in metals. 

In this paper we adopt a similar approach, and perform a stacking analysis of $\simeq 1000$ galaxies at $z>4$ for which NIRSpec prism observations have been obtained.  We use  the \emph{DAWN JWST Archive} \citep[DJA\footnote{\url{https://dawn-cph.github.io/dja} }; see][]{Heintz.2024} to obtain a large sample of galaxy spectra at $4<z<13$, which we then combine in subsets of $\simeq 200$ galaxies per bin, arriving at total effective integration times of several million seconds per spectral pixel.  This extremely deep spectroscopy reveals the UV line emission from the galaxies, which we then measure to explore astrophysical processes in concert with the optical lines.  We study emission lines of [\cII], \cIII], \cIV, \heII, \oIII], \nIII] and \nIV] in the UV, as well as many strong lines in the optical range.  We make a concerted effort to understand: (1.) the nebular gas temperatures; (2.) the associated nebular abundances of C, N, and O and variation in their patterns; (3.) the production efficiency of ionizing radiation; and (4.) the ionizing escape fraction.  With information on the stellar population age, we investigate all this as a function of stellar evolutionary stage.

The paper proceeds as follows.  
In Section~\ref{sect:data} we present the sample selection (\S\ref{sect:data:samp}), data processing and spectral coaddition (\S\ref{sect:data:stack}), and emission line measurement methods (\S\ref{sect:data:meas}).  
In Section~\ref{sect:results} we show the average spectra of high-$z$ galaxies, describe the main trends with star-formation evolutionary state, quantify the lines ratios and EWs, and contrast the stacked high-$z$ results with well-studied low-$z$ samples -- we do this qualitatively in \S\ref{sect:stackdescribe} and quantitatively in \S\ref{sect:lowzcomp}. 
In Section~\ref{sect:abund} we calculate electron temperatures using both the Balmer Jump (\S\ref{sect:bj}) and collisional lines of [\oIII] (\S\ref{sect:TeOIII}), estimate the abundances of oxygen, carbon, and nitrogen (\S\ref{sect:meth:abund}) and discuss the impact of our assumptions on these results (\S\ref{sect:testassumpt}). 
We describe measurements related to the production and escape of ionizing radiation (\fesclyc\ and \xiion) in Section~\ref{sect:res:ion}. 
In Section~\ref{sect:disc} we discuss how these observations inform our view of the ionizing conditions in early galaxies (\S\ref{sect:disc:ioncond}), gas temperatures (\S\ref{sect:disc:temp}), existing efforts to understand the buildup and chemical enrichment of galaxies (\S\ref{sect:disc:abunda}), and the emission of ionizing radiation (\S\ref{sect:disc:ionrad}).  
In Section~\ref{sect:conc} we present our concluding remarks.  Throughout we assume a cosmology of $\{H_0,\Omega_\mathrm{M},\Omega_\Lambda\} = \{70~\mathrm{km~s}^{-1}~\mathrm{Mpc}^{-1},0.3, 0.7\}$, and the AB magnitude system \citep{Oke1983}.

\section{Sample, Methods, and Measurements}\label{sect:data}

Here we describe the galaxy sample, processing methods and stacking, and measurements we make on the co-added spectra. 

\subsection{Sample Selection}\label{sect:data:samp}

\begin{figure*}
\includegraphics[width=18cm]{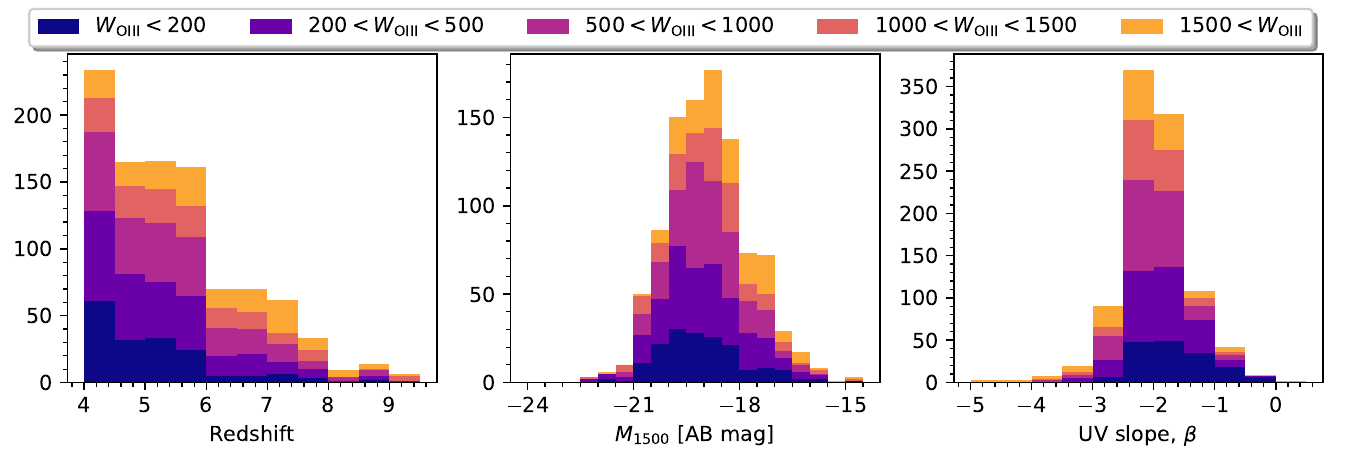}
\caption{Basic sample characteristics.  Histograms from left to right show redshift, UV absolute magnitude, and UV continuum slope $\beta$.  All are stacked by dependent variable \ewoIII, in a color coding that follows throughout the manuscript.}
\label{fig:sample}
\end{figure*}

The sample is drawn from the DJA archive, similar to \citet[][see also \citealt{Roberts-Borsani.2024} and \citealt{Katz.2024}]{Heintz.2024}.  We require a large number of galaxies and complete spectral coverage between \lya\ and \halpha, which is satisfied by the samples for which PRISM/CLEAR spectroscopy have been obtained.  The prism has a wavelength range of 0.6--5.3~$\mu$m, with a resolving power of $R\equiv \lambda/\Delta\lambda=35-300$, depending upon the wavelength. Our main aim is to obtain the deepest possible spectrum of the UV nebular lines, so we set a lower limit of $z\ge 4$, which shifts the \lya\ break and emission line into the wavelength coverage of NIRSpec/PRISM.  Wishing to study galaxies to the largest possible distance, we set no upper limit in redshift.  Each object in the DJA spectroscopic archive includes a \texttt{grade} flag, which identifies the quality of the spectrum; we take only spectra with \texttt{grade}\,$=3$, corresponding to the highest quality sources, although all spectra were also visually inspected. This yields a sample of 1115 galaxies.  They are drawn from the multitude of JWST spectroscopic campaigns from all cycles, including General Observing campaigns, Early Release Science, Guaranteed Time Observations, and Director's Discretionary Time, and combine all kinds of different selection techniques. 

From this we select only galaxies with PRISM/CLEAR observations, removing those with only observations with the higher-resolution gratings.  Some spectra show significant artifacts or corrupted measurements with extended spectral regions of the spectra being identically zero or not-a-number in the spectra.  We reject any object with $>50$\% corrupted pixels in the restframe wavelength range between 1200 and 1900\AA\ (\lya\ to \cIII]), assuming the default redshifts assigned in the DJA archive. 

\subsection{Continuum Modeling and Stacking}\label{sect:data:stack}

Our goal is to study line emission, so the stellar continuum must either be removed by subtraction or normalized.  We make use of both procedures in this paper: subtraction of continuum conserves line fluxes, allowing them to be measured and line ratios calculated; normalization of the continuum conserves equivalent width of each line.  The best approach would be to model the stellar continuum with population synthesis models \citep[e.g][]{Hayes.2023lya}, but these individual spectra have neither the resolution nor signal-to-noise ratio (S/N) to make such fitting meaningful.  We therefore adopt simple polynomial representations of the continuum.  Such continuous functions will not capture spectral discontinuities such as the Balmer/4000\AA\ break, so we perform this functional fitting over two separate wavelength regions for each object: the UV (at $\lambda_\mathrm{rest}<3500$\AA\ to avoid the Balmer break) and optical (at $\lambda_\mathrm{rest}>3800$\AA\ to avoid the [\oII] emission line). 

We mask regions covered by the main expected strong lines in the UV and optical (see Section~\ref{sect:data:meas} for the lists), in windows of $\pm 1500$~\kms\ (I.e. the average spectral resolution, implicitly assuming lines are unresolved).  We then fit third-order polynomial functions to the UV and optical regions independently.  We also fit the UV range (1500 to 2200\,\AA) with a power-law, to recover the UV continuum slope, $\beta$ and use it as a test parameter for differential comparisons\footnote{The conventional description of the UV color, given in wavelength flux densities, is $f_\lambda \propto \lambda^\beta$}. 

To verify, and in some cases improve, the redshift precision for the stacking we make a first round of fits to emission lines in each spectrum.  We assume that most of the important information for redshifts will be present in the strong optical emission lines of [\oII], \hbeta, [\oIII], and \halpha, and that most of these lines could be visible for individual galaxies.  We perform a simultaneous fit to these emission lines using the same engine as used in \citet{Hayes.2023lya}.  Fixing the redshifts of all features to be the same, we fit single Gaussian functions for [\oII], \hbeta, and \halpha\ (ignoring possible contamination from the adjacent [\nII] doublet), and a double Gaussian to the [\oIII] feature in which the 4960\AA\ emission line is fixed to be 1/3 the strength of the 5008\AA\ line.  The amplitude for each line is its only individual free parameter and, thus, the four features are fit to provide a single redshift.  We note that the \halpha\ line is lost from the red end of NIRSpec at $z>6.8$ and [\oIII]+\hbeta\ are lost at $z>9.5$.  We deem it unreliable to base redshifts purely upon [\oII] line, so limit our study to galaxies where [\oIII], \hbeta, and [\oII] are available, giving us a de facto cap of $z<9.5$. We declare a fit to be `successful' if we obtain $S/N>5$ measurements in any of the features, which provides the effective selection function of our study within the DJA archive. All fits were inspected by eye.  Removing a handful of galaxies for which this line fitting fails, we are left with a sample of 1035 galaxies. We show the basic properties of this sample in Figure~\ref{fig:sample} and provide more details  in the coming sections. 

With redshifts based upon spectral lines, we shift each spectrum into the restframe, and resample them onto a common rest wavelength grid spaced at 5\AA.  We transform flux density to luminosity density $L_\lambda$, by multiplying by $4\pi d_\mathrm{L}^2 (1+z)$, where $d_\mathrm{L}$ is the luminosity distance.   We produce four grids of the restframe spectra: (1.) those with no normalization so we can investigate the properties of the spectra in absolute luminosities; (2.) those normalized by the monochromatic continuum luminosity at 3500\AA, so we can investigate changes in continuum colors; (3.) those with the local continuum subtracted at each wavelength (using the fitted polynomial), so we can measure emission lines luminosities in their absolute units; and (4.) those with the variable continuum normalized (again at each wavelength using the same polynomial fit), so that equivalent width can be derived.  We compute uncertainties on the stacked spectra using a bootstrap-with-replacement simulation with 1000 realizations. 

\begin{deluxetable}{cccc}[htb!]
\tablecaption{Descriptive properties of galaxies that comprise the five stacks\label{tab:sampdesc}}
\tablehead{
\colhead{stack \#}  &  \colhead{$W_{\mathrm{[OIII]}}$ (\AA)}  &  \colhead{$M_{\mathrm{1500}}$}  &  \colhead{$\beta$}  }
\startdata
     $1$  &     $ 116^{+  48}_{-  56}$  &     $-19.07^{+0.81}_{-0.76}$  &     $-1.68^{+0.43}_{-0.40}$  \\ 
     $2$  &     $ 348^{+  83}_{-  90}$  &     $-19.07^{+0.82}_{-0.75}$  &     $-1.92^{+0.35}_{-0.25}$  \\ 
     $3$  &     $ 699^{+ 151}_{- 107}$  &     $-18.99^{+0.74}_{-0.56}$  &     $-2.05^{+0.27}_{-0.30}$  \\ 
     $4$  &     $1223^{+ 117}_{- 107}$  &     $-18.82^{+0.63}_{-0.93}$  &     $-2.09^{+0.30}_{-0.23}$  \\ 
     $5$  &     $2148^{+3711}_{- 411}$  &     $-18.50^{+0.89}_{-0.63}$  &     $-2.20^{+0.35}_{-0.35}$  \\ 
\enddata
\end{deluxetable}

In this procedure we have recorded a number of quantities.  We store \Muv\ (1500\AA) and $\beta$ (over the 1500--2200\AA\ range) from the power-law fitting to the UV portion of the spectrum, and the fluxes and EWs in [\oII], \hbeta, [\oIII], and \halpha. We use these tabulated variables as vectors along which we sub-divide the sample into bins, and stack each of the grids of spectra.  Thus we can investigate changes in the average spectra as a function of these properties.  In this paper we use the optical [\oIII]$\lambda5008$ equivalent width (\ewoIII) as our independent variable, because it is the strongest spectral feature that we have in the optical spectrum. \ewoIII\ closely traces the youth of the ongoing star formation episode as at most metallicities it is primarily dependent upon the number of ionizing photons produced per unit time\footnote{This relationship must break at very low metallicities, where the [\oIII] emission becomes sub-dominant to \hbeta, as has been potentially identified in very faint galaxies by \citet{Endsley.2024}, although see also \citet{Begley.2024}.  In the \citet{Endsley.2024} sample, this effect becomes most prominent in their \Muv\,$\simeq-17.5$, of which there are very few such galaxies in our sample (Figure~\ref{fig:sample}); while this must affect our work to some degree, we expect the impact to be minor (especially given the median stacking performed).}.  We divide the galaxies into five approximately equally spaced bins, at boundaries of 200, 500, 1000, 1500\,\AA, with total integration times in each stack reaching several million seconds.  The basic properties of these individual stacks are shown in Table~\ref{tab:sampdesc}.

\subsection{Emission Line Measurements}\label{sect:data:meas}

\begin{deluxetable}{ccl}[htb!]
\tablehead{
\colhead{Vacuum}     & \colhead{Target}  & \colhead{Comments} \\
\colhead{wavelength} & \colhead{line(s)} & \\
\colhead{(\AA)} &  &  }
\startdata
1487 & \nIV]~$\lambda\lambda1483,86$     & blended doublet \\ 
1550 & \cIV~$\lambda\lambda1548,51$      & blended resonance doublet \\ 
1640 & \heII~$\lambda1640$               & -- \\ 
1663 & \oIII]~$\lambda1661$+             & blended pair, marginally   \\ 
     & \oIII]~$\lambda1666$              & ~~resolved from \heII \\ 
1749 & \nIII]~$\lambda\lambda$1747, 1749    & blended quintuplet \\ 
     & 1750, 1752, 1754                  & \\ 
1908 & [\cIII]~$\lambda1907$             & blended doublet \\  
     & \cIII]~$\lambda1909$              &  \\  
2326 & \cII]~$\lambda\lambda2323, 2324$  & blended quintuplet \\  
     & 2325, 2326, 2328                  &   \\  
2880 & \mgII~$\lambda\lambda2796,2803$   & blended resonance doublet \\ 
\hline
3727 & [\oII]~$\lambda\lambda 3727,29$ & blended doublet \\
3869 & [\neIII]~$\lambda3869$ & -- \\
3889  & \heI~$\lambda3889$ & marginally resolved \\
 &  & ~~from [\neIII] \\
3970 & [\neIII]~$\lambda3968$+\hepsilon & blended pair, unresolved\\
4102 & \hdelta  & -- \\  
4341 & \hgamma  & -- \\  
4363 & [\oIII]~$\lambda4363$  & marginally resolved \\  
 &  & ~~from \hgamma \\  
4687 & \heII~$\lambda$4687 & -- \\ 
4862 & \hbeta  & -- \\ 
4960 & [\oIII]~$\lambda4960$  & -- \\ 
5008 & [\oIII]~$\lambda5008$  & 4960 doublet companion \\ 
5877 & \heI~$\lambda5877$  & -- \\ 
6302 & [\oI]~$6302$ & -- \\ 
6564 & \halpha~$\lambda6564$+ & blended complex, \\ 
 & [\nII]~$\lambda\lambda6549,85$ & ~~unresolved \\ 
6724 & [\sII]~$\lambda\lambda6717,31$  & blended doublet\\  
\enddata
\caption{Summary of the UV and optical emission lines considered in this work.  Describes the wavelengths used in the fitting modules and what features they measure.}
\label{tab:linesfit}
\end{deluxetable}

For each set of stacked spectra, we measure the luminosities (in the continuum-subtracted spectra) and EWs (in the normalized spectra) for a large number of UV and optical lines. Some lines, particularly a number of UV doublets, are blended at the resolution of NIRSpec's PRISM, and cannot be measured independently. We list the lines we fit in Table~\ref{tab:linesfit}, providing comments where appropriate. 

\begin{figure*}
\includegraphics[width=0.95\textwidth]{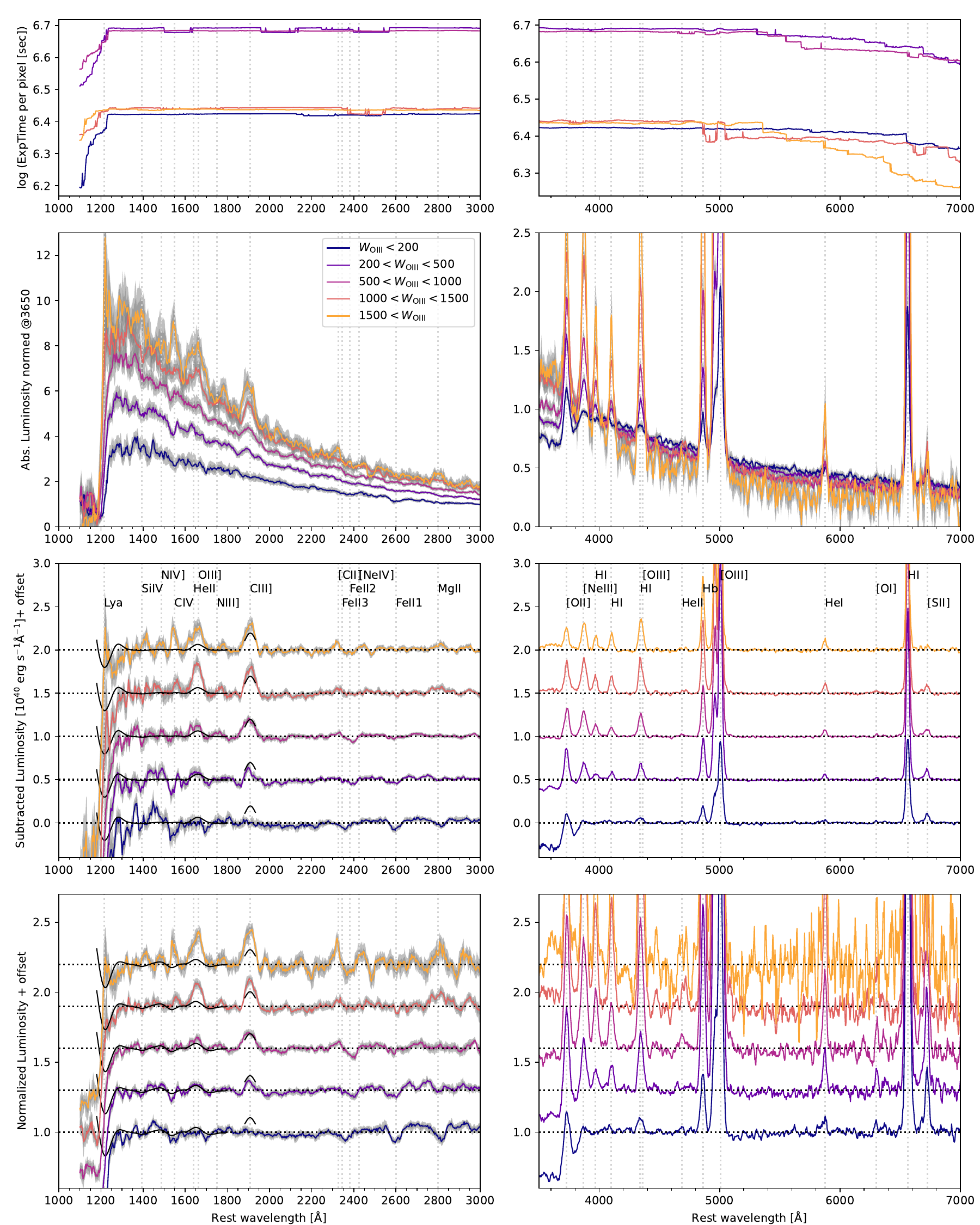}
\caption{Stacked spectra, ordered by \ewoIII.  Ultraviolet spectra are shown to the left and optical spectra to the right.  Upper panels show the logarithm of the total integration time per spectral pixel.  The second row shows the combination of spectra normalized by a single luminosity density, measured at 3500\AA. The third row shows stacks where the continuum has been subtracted from the individual spectra at each wavelength before coaddition.  Lower panels are the same except the continuum has been normalized to one at each wavelength.  Spectra in the lower four panels are shown with an arbitrary offset for clarity, with \ewoIII\ increasing from bottom to top.  Gray shaded regions surrounding each line indicate the 16--84th percentile ranges of a bootstrap simulation.  The black spectra in the UV panels show the stacked spectra of the CLASSY sample with \ewoIII~$> 1000$\AA, for which the luminosity has been multiplied by a factor of 10. }
\label{fig:spec_wOIII}
\end{figure*}

In the UV, we are unable to deblend doublets of \cIV$\lambda\lambda 1548,1551$, \oIII]$\lambda\lambda 1661,1666$, \cIII]$\lambda\lambda 1907,1909$,  \mgII$\lambda\lambda 2796,2803$, or the \nIII] multiplet (1747, 1749, 1750, 1752, and 1754\AA). Importantly, however, we are able to deblend the \heII$\lambda1640$ line from the \oIII] doublet -- these lines are separated by restframe $\simeq 20$\,\AA, which corresponds to marginally more than one resolution element.  This can be seen by contrasting the \heII+[\oIII] complex with the \cIV\ doublet in the lower left panel of Figure~\ref{fig:spec_wOIII}: the difference in spectral resolution is small across this wavelength range, but the \heII+[\oIII] features are clearly broader than \cIV, by a factor of around 2.  We impose a fixed wavelength separation between \heII\ and the \oIII] blend (see Table~\ref{tab:linesfit}), constrain the FWHM to be equal, and simultaneously fit for the amplitude of two Gaussian components.  In the optical, we are  unable to deblend the doublets of [\oII]$\lambda\lambda 3726,3729$ and [\sII]$\lambda\lambda 6717,6731$, as well as blends of [\nII]$\lambda 6584$+\halpha, and [\neIII]$\lambda3968$+\hepsilon.  We fit all these blends with a single emission line.  The [\oIII]$\lambda\lambda$4960,5008 lines are both resolved from each other and clearly separated from \hbeta, as well as [\neIII]$\lambda3869$ from \heI$\lambda3889$ and \hgamma\ from [\oIII]$\lambda4363$.  See Table~\ref{tab:linesfit} for a complete list.

We fit individual emission lines using single Gaussian profiles and applying the \textsc{lmfit} minimization algorithm \citep{lmfit.2016} to both continuum-subtracted (to get absolute line luminosities) and continuum-normalized \ewoIII-stacked spectra (for EWs). Free parameters are the line amplitude, the central rest-frame wavelength (in \AA, see priors in Table~\ref{tab:linesfit}) and the width of each line (in \AA).  We also record the luminosity and EW of each line by numerical integration, using a 5,000~\kms\ window.

The amplitude is constrained to be positive ($A_i \geq 0$) for all lines except for \cIV\ and \mgII, for which we also allow for negative values ($A_i < 0$) in order to account for stellar and ISM absorption components. Close-by emission lines within blended doublets or pairs are fitted simultaneously, and the wavelength separation between them is fixed to the value given by the rest-frame wavelength of the constituents. Close doublets and multiplets are fitted with single component Gaussians at the average wavelength -- see Table~\ref{tab:linesfit} for details.  

Errors on the line parameters were derived using a Monte-Carlo approach, perturbing the value of every spectral pixel by a Gaussian distribution whose mean is zero and standard deviation is the $1\sigma$ error coming from the bootstrap simulations, then re-fitting each line over $\times500$ iterations per stacked spectrum. The resulting line luminosities and EWs are presented in Sect.\ \ref{sect:stackdescribe}, and discussed in the context of this paper. 

\begin{figure}
\centering
\includegraphics[width=0.9\columnwidth]{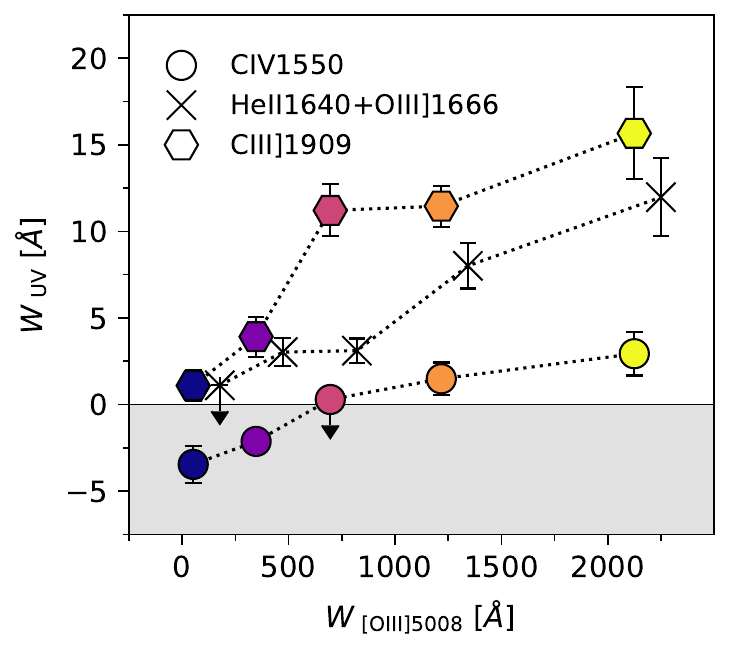}
\caption{Evolution of the EW (in \AA) of different UV features as a function of \ewoIII\ in the stacked spectra. \cIV$~\lambda\lambda1548,51$ transitions from absorption (stellar P-Cygni+ISM) to emission (nebular dominated). The strength of \heII$~\lambda1640$, \oIII]$~\lambda\lambda1661,66$, and \cIII]$~\lambda\lambda1907,09$ increase monotonically with \ewoIII, e.g., up to 16\AA\ in the case of \cIII].}
\label{fig:wUV_wOIII}
\end{figure}

\section{Results I: The Average Spectrum of Galaxies at Redshift Beyond 4 }\label{sect:results}

In Figure~\ref{fig:spec_wOIII} we show the average spectra of the galaxies in our sample, when binned in five groups sorted along a vector of \ewoIII.  \ewoIII\ probes, on average, the youth of the star-forming episode by tracing the ionizing photon production rate, modulated by the ionization parameter and the nebular oxygen abundance. The highest values are found among galaxies hosting very young, intense starbursts with weak stellar continua, and $12+\log(\mathrm{O/H}) \simeq 8$ (e.g. \citealt{Izotov.2023} and the sample of \citealt{Hayes.2023lya}, at least at low-$z$).  [\oIII] is often the brightest line in the optical spectrum of starbursts, and visible with NIRSpec/PRISM out to $z=9.5$.  These bins have median values of \ewoIII\ ranging from 120\AA\ for the lowest EW bin and 2200\AA\ for the highest (See Table~\ref{tab:sampdesc}).

\begin{deluxetable*}{ccccccc}[htb!]
\tablecaption{UV line luminosities measured in each stack in units of $10^{40}$\,erg\,s$^{-1}$.\label{tab:fluxuv}}
\tablehead{
\colhead{stack \#}  &  \colhead{C{\sc iv}/1550}  &  \colhead{He{\sc ii}/1640}  &  \colhead{[O{\sc iii}]/1666}  &  \colhead{C{\sc iii}]/1909}  &  \colhead{[C{\sc ii}]/2326}  &  \colhead{Mg{\sc ii}/2800}  
}
\startdata
      1  &  $-7.81 \pm 2.58$  &  $0.87 \pm 2.53$  &  $0.93 \pm 2.64$  &  $1.89 \pm 1.73$  &  $-4.10 \pm 1.60$  &  $-4.40 \pm 1.09$  \\
      2  &  $-5.20 \pm 1.33$  &  $3.12 \pm 1.10$  &  $3.04 \pm 1.05$  &  $6.49 \pm 1.73$  &  $5.03 \pm 0.91$  &  $-1.63 \pm 0.58$  \\
      3  &  $0.71 \pm 1.90$  &  $2.57 \pm 0.88$  &  $3.31 \pm 1.03$  &  $18.76 \pm 2.11$  &  $7.56 \pm 1.72$  &  $0.22 \pm 1.78$  \\
      4  &  $3.09 \pm 2.28$  &  $7.57 \pm 2.15$  &  $8.90 \pm 2.77$  &  $17.93 \pm 2.44$  &  $1.85 \pm 1.06$  &  $10.74 \pm 2.34$  \\
      5  &  $5.18 \pm 1.55$  &  $4.81 \pm 2.81$  &  $10.75 \pm 2.38$  &  $17.25 \pm 2.69$  &  $4.41 \pm 1.44$  &  $2.99 \pm 2.08$  \\
\enddata
\end{deluxetable*}

\begin{deluxetable*}{ccccccccc}[htb!]
\tablecaption{Optical line luminosities measured in each stack in units of $10^{40}$\,erg\,s$^{-1}$.\label{tab:fluxop}}
\tablehead{
\colhead{stack \#}  &  \colhead{[O{\sc ii}]/3728}  &  \colhead{[O{\sc iii}]/4364}  &  \colhead{He{\sc ii}/4687}  &  \colhead{H{\sc i}/4862}  &  \colhead{[O{\sc iii}]/5008}  &  \colhead{H{\sc i}/6564}  &  \colhead{[S{\sc ii}]/6718}  &  \colhead{[S{\sc ii}]/6732}  
}
\startdata
      1  &  $6.98 \pm 0.73$  &  $1.99 \pm 3.04$  &  $0.82 \pm 0.47$  &  $6.62 \pm 0.86$  &  $37.14 \pm 3.13$  &  $34.62 \pm 3.34$  &  $3.13 \pm 0.96$  &  $0.66 \pm 0.59$  \\
      2  &  $16.33 \pm 2.08$  &  $3.40 \pm 0.89$  &  $0.00 \pm 0.48$  &  $16.81 \pm 1.33$  &  $88.95 \pm 7.52$  &  $58.98 \pm 5.38$  &  $4.04 \pm 0.42$  &  $0.00 \pm 0.00$  \\
      3  &  $15.42 \pm 1.76$  &  $1.49 \pm 1.57$  &  $0.83 \pm 0.47$  &  $19.83 \pm 0.97$  &  $120.66 \pm 6.61$  &  $71.59 \pm 3.36$  &  $5.49 \pm 1.57$  &  $0.00 \pm 0.00$  \\
      4  &  $14.06 \pm 1.47$  &  $7.72 \pm 1.13$  &  $1.16 \pm 0.49$  &  $24.25 \pm 1.94$  &  $151.44 \pm 16.98$  &  $89.11 \pm 11.05$  &  $3.33 \pm 2.63$  &  $0.00 \pm 0.72$  \\
      5  &  $8.59 \pm 1.34$  &  $4.30 \pm 2.78$  &  $10.43 \pm 1.52$  &  $23.22 \pm 2.00$  &  $122.99 \pm 10.32$  &  $68.94 \pm 7.55$  &  $0.23 \pm 3.16$  &  $0.76 \pm 0.32$  \\
\enddata
\end{deluxetable*}

\begin{deluxetable*}{ccccccc}[htb!]
\tablecaption{UV line equivalent widths measured in each stack in units of \AA.\label{tab:ewuv}}
\tablehead{
\colhead{stack \#}  &  \colhead{C{\sc iv}/1550}  &  \colhead{He{\sc ii}/1640}  &  \colhead{[O{\sc iii}]/1666}  &  \colhead{C{\sc iii}]/1909}  &  \colhead{[C{\sc ii}]/2326}  &  \colhead{Mg{\sc ii}/2800}  
}
\startdata
      1  &  $-3.46 \pm 1.09$  &  $1.88 \pm 0.76$  &  $-0.77 \pm 0.84$  &  $1.09 \pm 0.89$  &  $-3.99 \pm 1.06$  &  $-5.63 \pm 1.28$  \\
      2  &  $-2.13 \pm 0.47$  &  $1.54 \pm 0.51$  &  $1.48 \pm 0.63$  &  $3.92 \pm 1.16$  &  $4.23 \pm 0.80$  &  $-2.98 \pm 1.11$  \\
      3  &  $0.29 \pm 0.88$  &  $1.20 \pm 0.50$  &  $1.91 \pm 0.49$  &  $11.21 \pm 1.50$  &  $7.59 \pm 1.48$  &  $0.23 \pm 2.03$  \\
      4  &  $1.49 \pm 0.94$  &  $3.87 \pm 0.79$  &  $4.14 \pm 1.04$  &  $11.46 \pm 1.19$  &  $1.68 \pm 1.13$  &  $13.78 \pm 2.99$  \\
      5  &  $2.93 \pm 1.26$  &  $4.06 \pm 1.53$  &  $7.92 \pm 1.67$  &  $15.67 \pm 2.66$  &  $7.26 \pm 1.75$  &  $8.74 \pm 3.19$  \\
\enddata
\end{deluxetable*}

\begin{deluxetable*}{ccccccccc}[htb!]
\tablecaption{Optical line equivalent widths measured in each stack in units of \AA.\label{tab:ewop}}
\tablehead{
\colhead{stack \#}  &  \colhead{[O{\sc ii}]/3728}  &  \colhead{[O{\sc iii}]/4364}  &  \colhead{He{\sc ii}/4687}  &  \colhead{H{\sc i}/4862}  &  \colhead{[O{\sc iii}]/5008}  &  \colhead{H{\sc i}/6564}  &  \colhead{[S{\sc ii}]/6718}  &  \colhead{[S{\sc ii}]/6732}  
}
\startdata
      1  &  $9.62 \pm 0.67$  &  $4.67 \pm 4.63$  &  $1.08 \pm 0.68$  &  $16.91 \pm 1.56$  &  $94.56 \pm 4.82$  &  $147.33 \pm 7.21$  &  $12.35 \pm 2.11$  &  $1.31 \pm 1.62$  \\
      2  &  $34.93 \pm 1.54$  &  $5.46 \pm 4.68$  &  $0.00 \pm 0.00$  &  $51.93 \pm 1.30$  &  $303.51 \pm 6.60$  &  $353.50 \pm 7.84$  &  $23.23 \pm 1.09$  &  $0.00 \pm 0.17$  \\
      3  &  $46.36 \pm 2.97$  &  $0.00 \pm 2.41$  &  $3.74 \pm 4.19$  &  $108.73 \pm 2.21$  &  $656.00 \pm 8.10$  &  $733.61 \pm 13.64$  &  $37.60 \pm 4.70$  &  $0.00 \pm 0.00$  \\
      4  &  $68.09 \pm 4.75$  &  $37.78 \pm 3.71$  &  $6.42 \pm 0.48$  &  $203.43 \pm 5.03$  &  $1171.92 \pm 17.74$  &  $1159.58 \pm 46.06$  &  $8.04 \pm 3.70$  &  $31.75 \pm 4.35$  \\
      5  &  $76.41 \pm 4.86$  &  $60.36 \pm 10.87$  &  $20.24 \pm 3.74$  &  $376.71 \pm 17.65$  &  $2202.32 \pm 65.96$  &  $2790.32 \pm 233.83$  &  $13.10 \pm 14.74$  &  $21.60 \pm 7.08$  \\
\enddata
\end{deluxetable*}

The uppermost plots in Figure~\ref{fig:spec_wOIII} show the total integration time per spectral pixel, obtained by summing the integration times of all the spectra that enter the stack. These range from 2.5 to 5 million seconds, and are reasonably flat at $\lambda_\mathrm{rest}\gtrsim 1200$\AA, after \lya\ shifts into the prism coverage at $z\simeq 4$. At $\lambda_\mathrm{rest}\gtrsim 5000$\AA\ the exposure times per pixel start to decrease: we set our upper redshift limit of 9.5 to retain [\oIII]5008, and hence the drop begins shortly after this rest wavelength.  Curves then decline to longer wavelengths as redder radiation is lost down to $z\simeq 6.7$. The [\sII]$\lambda\lambda6717,31$ doublet is the reddest feature used in this work.  

\subsection{Evolution of the Average UV Spectrum with \ewoIII}\label{sect:stackdescribe}

The UV spectral range contains permitted recombination lines in emission, and various semi-forbidden and forbidden lines excited by collisions.  We draw attention to several key features in these UV spectra, and how they change as a function of \ewoIII\ in the optical.  Most notably: \\
-- Emission lines of \cIV$\lambda 1550$, the blend of \heII$\lambda 1640$ and \oIII]$\lambda\lambda 1661,66$, and \cIII]$\lambda 1909$ are all particularly clear, and seen in emission in the stacks of highest \ewoIII\ (e.g. yellow curve in the lower left panel of Figure~\ref{fig:spec_wOIII}).  The higher excitation energies of these UV lines compared to most optical lines observed at high-$z$ makes them detectable only at the high temperatures associated with low gas-phase metallicities \citep[e.g.][]{Mingozzi.2022}. In contrast, none of these features are seen in emission for the lowest \ewoIII\ bins (dark purple curve in the same figure).  Together with optical emission lines, these features can be used to estimate the electron temperatures and abundances of oxygen and carbon in the interstellar medium (see Section~\ref{sect:disc:abunda}).  \\
-- The \mgII$\lambda\lambda 2796$,2803\AA\ resonance doublet (blended) is also seen in emission for the highest EWs of [\oIII].  However, unlike the forbidden lines mentioned above, \mgII\ shifts to absorption for the less intensely star forming stacks, as it is scattered in the ISM.  In combination with optical line emission, \mgII\ can be used to infer ISM conditions pertinent to the escape of ionizing radiation \citep[e.g.][see Section~\ref{sect:disc:ionrad}]{Henry.2018,Xu.2022.lzlcs,Leclercq.2024}.  \\
-- Entirely analogous behavior is seen for the resonance doublet of \cIV$\lambda\lambda 1548$,1551\AA, which is also a ground-state resonance doublet and can form in both absorption and emission.  Among other things, \cIV\ has been suggested as a useful indirect diagnostic of LyC emission \citep{Schaerer.2022a.LycCIV}.  \\
-- [\cII] multiplet near $\lambda=2326$\AA\ is clearly visible in the high EW stacks.  This line has been commonly observed in spectra of various classes of AGN \citep{Best.2000,Humphrey.2014}, could be a good metallicity diagnostic \citep{Kewley.2019} and can provide ionization correction factors in the derivation of the carbon abundance. \\
-- There are detections of both the \nIV]$\lambda \lambda1483,1486$ and \nIII]$\lambda 1750$ multiplet in the highest \ewoIII\ stacks. These can provide estimates of the nitrogen abundance which in turn can point towards the presence of exotic stellar populations.  We discuss this in more detail in Section ~\ref{sect:disc:abunda}. \\
-- There is very clear evidence of both `normal' (red) and inverted (blue) Balmer jumps (see the lower right panel of Figure~\ref{fig:spec_wOIII}).  The red stellar jump is typically an indicator of stellar populations with ages $\sim 100$~Myr, while the blue nebular discontinuity encodes the temperature in the \hII\ gas without the bias towards hotter regions that is implicit in temperature diagnostics based upon collisionally excited lines.  The very fact that we can see the inverted jump indicates that the nebular continuum must contribute a significant fraction of the light in the $\lambda\sim 3650$\AA\ region \citep[e.g.][]{Katz.2024,Langeroodi.2024}. \\
-- The UV continuum shape becomes bluer for the highest \ewoIII\ galaxies, as seen in the second-left panel of Figure~\ref{fig:spec_wOIII}.  This is likely to be due to the stellar populations required to doubly ionize oxygen (driving the \ewoIII\ selection) being hotter and bluer in the non-ionizing UV. The UV slope is typically a combined indicator of the youth of the starburst, stellar metallicity, and dust reddening, as well as a signpost of possible LyC emission \citep{Chisholm.2022}. 

We also compare the stacked NIRSpec spectra with spectra obtained in the low-$z$ universe, for which we adopt the \emph{COS Legacy Archive Spectroscopic SurveY} (CLASSY; \citealt{Berg.2022, James.2022}).  The CLASSY sample comprises 45 vigorously star forming galaxies at $z<0.18$, for which deep and almost-homogeneous spectroscopy has been obtained with HST/COS in all of the medium resolution gratings. The primary selection criteria of CLASSY extracted a sample comparable in various characteristics (SFR, stellar mass, size, oxygen abundance, etc.) to those identified at redshifts above 6 \citep[see][for details]{Berg.2022}.  We treat these spectra similarly to those of the NIRSpec sample, first removing the continuum through subtraction and normalization, then smoothing them to the resolving power of NIRSpec appropriate for the wavelength. We then stack all the CLASSY sources with \ewoIII\,$>1000$\,\AA, which we plot as the  black lines in the lower left panels of Figure~\ref{fig:spec_wOIII}.  We note that CLASSY's COS spectra have been observed in the G185M grating to measure the \cIII]1909 line, but this grating has a very narrow spectral bandwidth which leads to a gap just bluewards of this line before the G160M grating.  We have no UV spectroscopy at redder wavelengths, so the [\cII] multiplet and \mgII\ emission are unknown in CLASSY for the moment. 

In the continuum-subtracted spectra, where the comparison is made directly in luminosity space, we have multiplied the CLASSY spectra by a factor of 10.  In the lower left plot we make the same comparison in normalized space.  Several similarities are visible between the CLASSY and NIRSpec spectra, but also a number of differences.  For example, attending to the \cIII]1909 emission line in the normalized spectra (lower left): this is stronger in the CLASSY stack than in high-$z$ galaxies with \ewoIII\,$<200$\,\AA.  However the EW of this line grows with increasing \ewoIII\ and roughly matches the CLASSY stack by \ewoIII\,$\simeq 500$~\AA, and becomes stronger still in the high-$z$ sample for the most intensely star forming galaxies -- the \ewoIII\,$\gtrsim 1500$\,\AA\ stack appears brighter in \cIII] by a factor of $\simeq 2.5$.  The \heII+\oIII] complex and \cIV\ line behave similarly to \cIII], broadly matching the CLASSY spectrum in the \ewoIII\,$\simeq 500$~\AA\ stack.    

\begin{figure*}
\includegraphics[height=4cm]{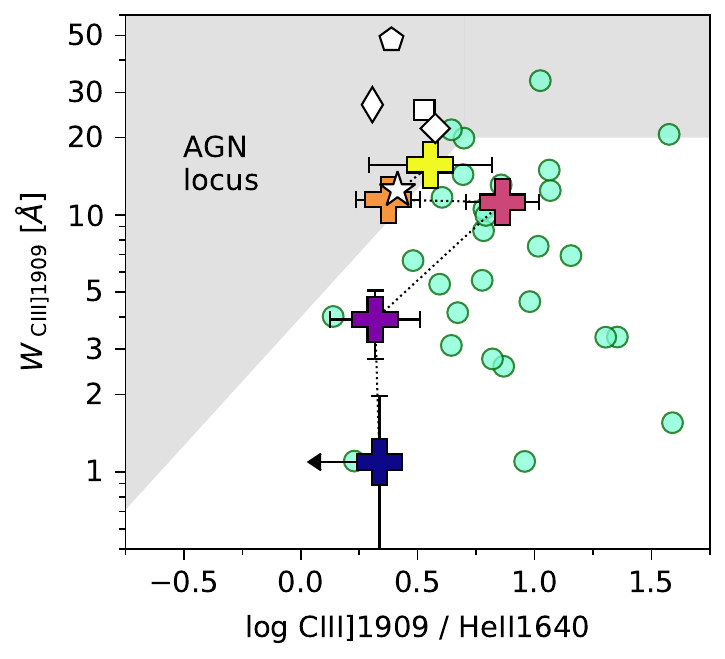}
\includegraphics[height=4cm]{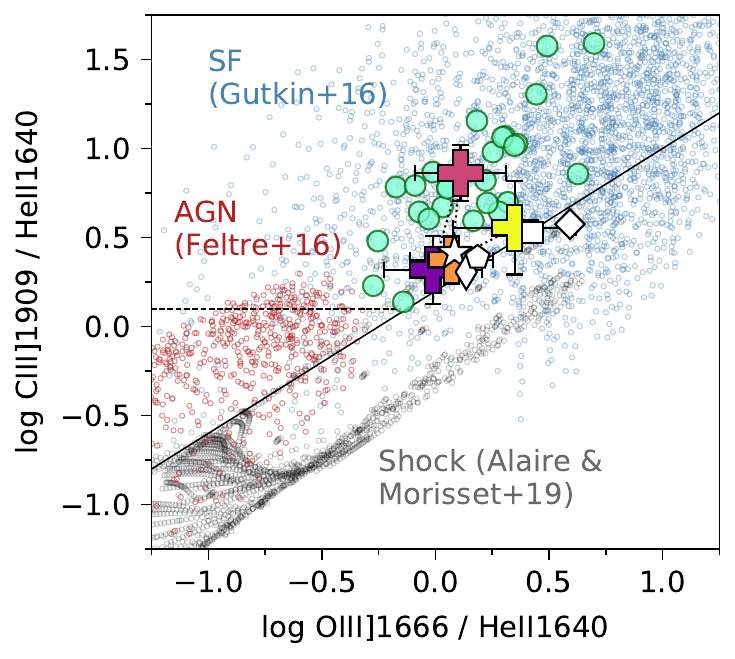}
\includegraphics[height=4cm]{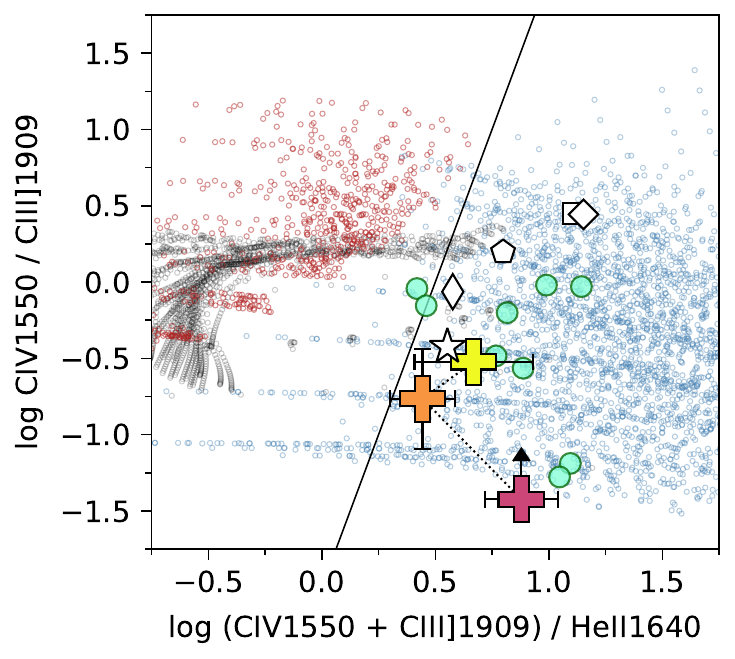}
\includegraphics[height=4cm]{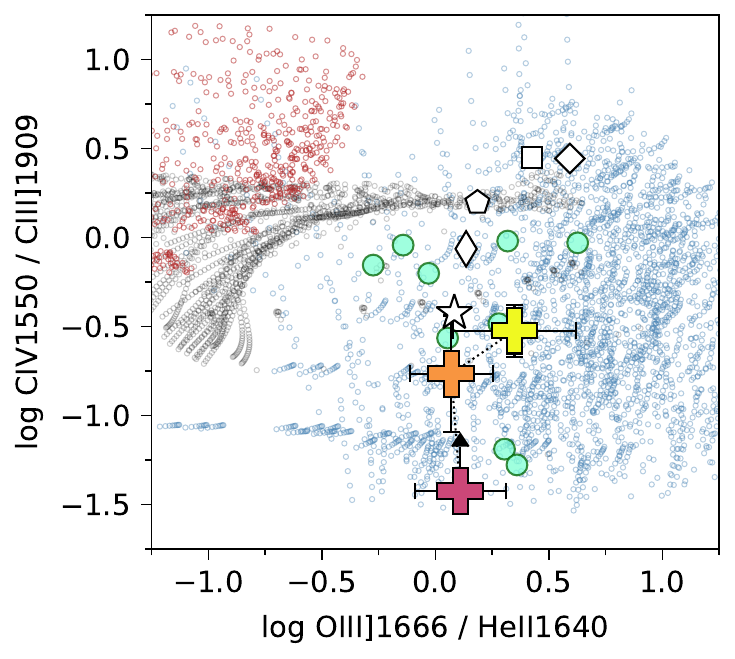}
\caption{Some diagnostic diagrams constructed from the UV line measurements. From left to right: equivalent width of \cIII] vs. \cIV/\cIII], \cIII]/\oIII] vs. \oIII]/\heII, \cIV/\cIII] vs. (\cIV+\cIII])/\heII\ and \cIV/\cIII] vs. \oIII]/\heII. Points with errorbars correspond in color to the \ewoIII-binned stacks presented in Figure~\ref{fig:spec_wOIII}, while green circles show the location of low-$z$ galaxies from the CLASSY survey, published in \citet{Mingozzi.2024}. The white star, square, thin diamond, pentagon and thick diamond show measurements for GNz11, GHZ12, CEERS-1019, GHZ9 and RXCJ2248-ID at high-$z$, respectively \citep{Bunker.2023, Castellano.2024, Marques-Chaves.2024, Napolitano.2024_GHZ9, Topping.2024}. The shaded region and solid line in the left-hand and middle-right plots display the AGN/SF demarcation line of \citet{Nakajima.2018}. Solid and dashed lines in the middle-left panel show the shock/AGN and AGN/SF from \citet{Mingozzi.2024}, respectively. Blue, red, and black dots in the background are the photoionization models for AGN, shock and star-formation in \citet{Feltre.2016}, \citet{AM.2019} and \citet{Gutkin.2016}, at sub-solar metallicity.}
\label{fig:uvbpt}
\end{figure*}

\begin{deluxetable*}{cccccccc}[htb!]
\tablecaption{List of commonly used line ratios.\label{tab:linerats}}
\tablehead{
\colhead{stack \#}  &  \colhead{H$\alpha$/H$\beta$}  &  \colhead{O$_{32}$}  &  \colhead{[S{\sc ii}](6717+31)/H$\alpha$}  &  \colhead{[O{\sc iii}]5008/H$\beta$}  &  \colhead{[O{\sc iii}]5008/(1661+66)}  &  \colhead{O{\sc iii}](1661+66)/H$\beta$}  &  \colhead{C{\sc iii}](1907+09)/H$\beta$}  
}
\startdata
      1  &  $5.23 \pm 0.85$  &  $5.32 \pm 0.72$  &  $0.11 \pm 0.03$  &  $5.61 \pm 0.87$  &  $>40.14$           &  $<0.14$  &  $0.29 \pm 0.26$  \\
      2  &  $3.51 \pm 0.42$  &  $5.45 \pm 0.83$  &  $0.07 \pm 0.01$  &  $5.29 \pm 0.61$  &  $29.30 \pm 10.43$  &  $0.18 \pm 0.06$  &  $0.39 \pm 0.11$  \\
      3  &  $3.61 \pm 0.25$  &  $7.83 \pm 0.99$  &  $0.08 \pm 0.02$  &  $6.08 \pm 0.45$  &  $36.41 \pm 11.45$  &  $0.17 \pm 0.05$  &  $0.95 \pm 0.12$  \\
      4  &  $3.67 \pm 0.54$  &  $10.77 \pm 1.65$  &  $0.04 \pm 0.03$  &  $6.24 \pm 0.86$  &  $17.02 \pm 5.63$  &  $0.37 \pm 0.12$  &  $0.74 \pm 0.12$  \\
      5  &  $2.97 \pm 0.41$  &  $14.32 \pm 2.53$  &  $0.01 \pm 0.05$  &  $5.30 \pm 0.64$  &  $11.44 \pm 2.71$  &  $0.46 \pm 0.11$  &  $0.74 \pm 0.13$  \\
\enddata
\end{deluxetable*}

\subsection{Quantitative Trends and Comparisons}\label{sect:lowzcomp}

We measure the strength of the UV emission lines, as described in Section~\ref{sect:data:meas}.  We show the evolution of some characteristic EWs in Figure~\ref{fig:wUV_wOIII}, and present the fluxes and equivalent widths in Tables~\ref{tab:fluxuv}, \ref{tab:fluxop}, \ref{tab:ewuv}, and \ref{tab:ewop}.  Several results are based upon line ratios, of which the major ratios used in this paper are listed in Table~\ref{tab:linerats}.   We first correct all emission lines for differential dust obscuration using the \citet{Reddy.2015,Reddy.2016} attenuation law, and dust reddening estimated from the \halpha/\hbeta\ line ratio assuming an intrinsic value of 2.86, which is appropriate for  $10^4$~K gas.

We assess their evolution on various diagnostic diagrams, and contrast the high-$z$ average spectra with other individual targets, in Figure~\ref{fig:uvbpt}.  This Figure shows diagrams of (a.) the \cIII]-EW vs. the \cIII]/\heII\ ratio (left-most), (b.) the \cIII]/\heII\ vs. the \oIII]/\heII\ ratio (center-left), (c.) the \cIV/\cIII] ratio vs. the (\cIII]+\cIV)/\heII\ ratio (center-right), and (d.)  \cIV/\cIII] ratio vs. \oIII]/\heII\ (right-most).  When galaxies are ordered by starburst intensity (increasing \ewoIII), it is clearly visible how the \cIII] EW also increases, rising from $\sim 1$\AA\ to 15\AA\ for the most intense starburst (e.g. central panel).  This evolution in EW is unsurprising, since these lines are, to first order, connected to the rate of ionizing photon production at energies above $\simeq 35$ and $\simeq 24$~eV, respectively, and naturally we expect their EWs to broadly correlate.  However, the two right-most plots show that the \cIV/\cIII] ratio also increases along the same sequence of \ewoIII, implying there is also an increase in the hardness of the ionizing spectrum (probed here by ionization edges at 45 and 24~eV).  

Light green points show the position of galaxies from the CLASSY sample \citep{Berg.2022,James.2022}, including only galaxies for which all these flux measurements can be made individually (i.e. showing no limits).  This paints a different picture to the qualitative discussion based on the stacked CLASSY spectra (Section~\ref{sect:stackdescribe}): the \cIV/\cIII] ratios of the stacked high-$z$ spectra reach values of $\simeq 0.4$, while the median of green points is higher, and reaches values of almost 1.   This happens despite the fact that both lines must be detected to appear on the diagram, and the real dispersion in the low-$z$ sample remains very large compared to the averages of the sample.  

AGN, star-forming, and shocked gas regions are overplotted from models of \citet{Feltre.2016}, \citet{Allen.2008}, and \citet{Gutkin.2016} in the line ratio diagrams, as well as demarcated from \citet{Nakajima.2018}.  It is clear that our stacks, as well as all the CLASSY galaxies, are well confined to the star-formation dominated regions in the line ratios and do not require additional sources of ionization. However, the two highest \ewoIII\ stacks, as well as a handful of CLASSY galaxies, do encroach into the AGN region when \ewcIII\ is used on one of the axes.  See Section~\ref{sect:disc} for more discussion on this.

\section{Results II: Electron Temperatures and Metal Abundances}\label{sect:abund}

The spectra show various nebular emission lines from the \opp\ ion, including the strong lines at 4960 and 5008\,\AA, the auroral line at 4363\,\AA, and a blend of the intercombination lines at 1661 and 1666\,\AA\ in the ultraviolet.  We also see a significant change in the amplitude of the Balmer Jump (BJ) near 3650\,\AA\ (Figure~\ref{fig:spec_wOIII}, lower panels), which we highlight in Figure~\ref{fig:bj}. The ratios of [\oIII]5008/1666 and [\oIII]5008/4363, as well as the amplitude of the Balmer Jump, all provide estimates of the characteristic kinetic energy of the electrons in the ionized gas, and therefore the electron temperature, \etemp.  Because most of the cooling in nebular gas occurs through collisions with ionized metals (primarily \op\ and \opp), this temperature is fundamentally influenced by the metal abundance.  We now explore each of these topics and measurements in turn.

\subsection{Electron Temperatures}\label{sect:temp}

\subsubsection{Balmer Jump}\label{sect:bj}

\begin{figure}
\centering
\includegraphics[width=0.9\columnwidth]{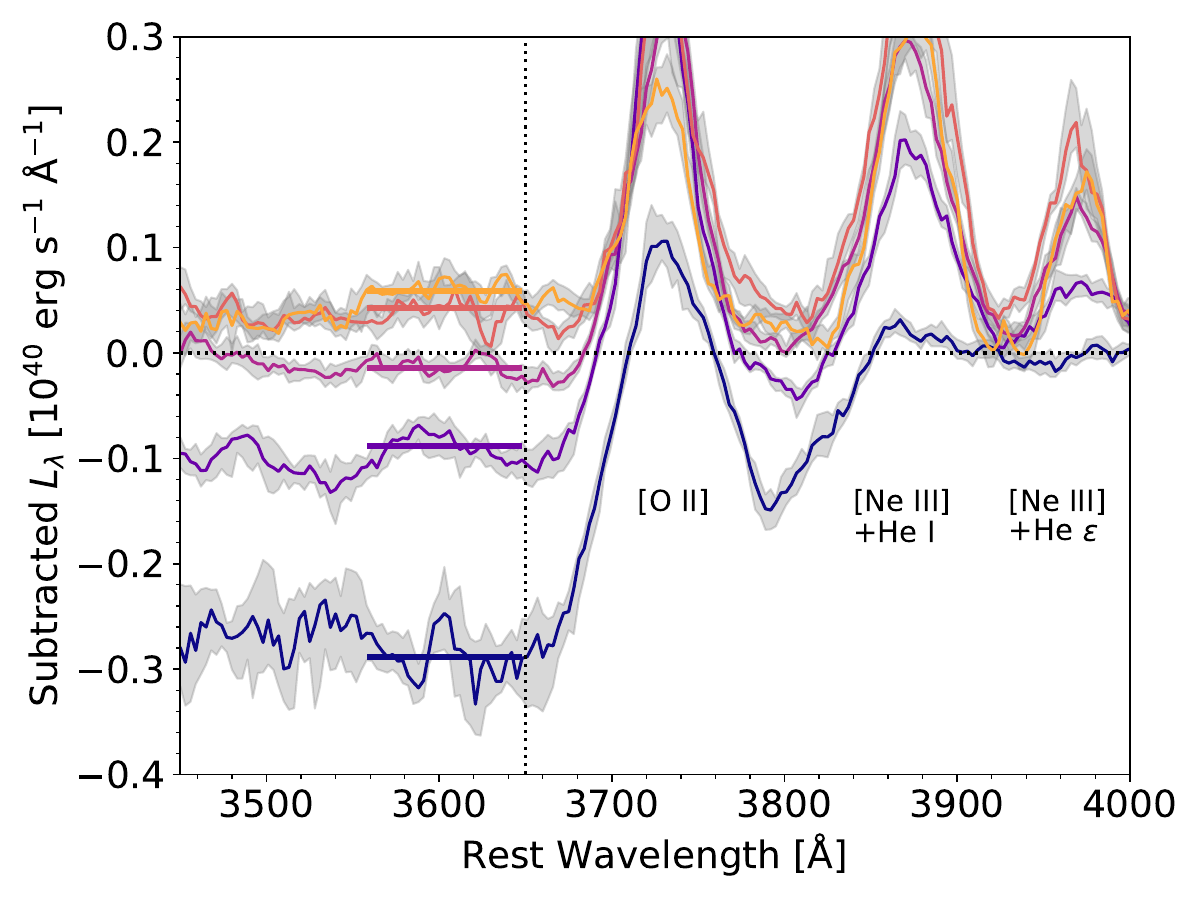}
\caption{Zoom of the Balmer Jump region of the spectra.  These spectra have been continuum subtracted using a  polynomial fit to the continuum at wavelengths longer than 3800\AA, which have been extrapolated to bluer wavelengths}.  Strong emission lines are labeled.  The color coding is the same as used in Figure~\ref{fig:spec_wOIII}.
\label{fig:bj}
\end{figure}

\begin{figure*}
\includegraphics[height=7.4cm]{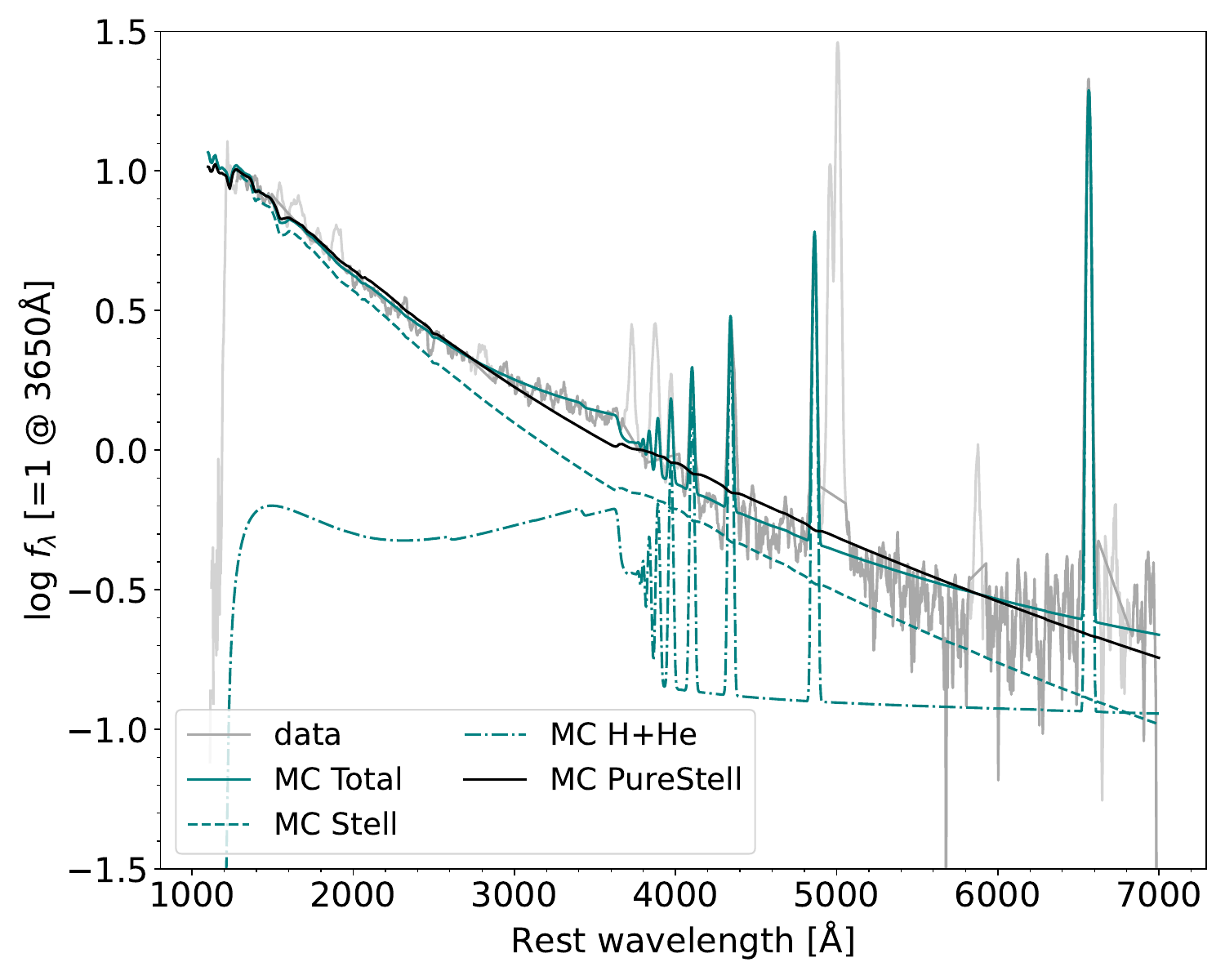}
\includegraphics[height=7.4cm]{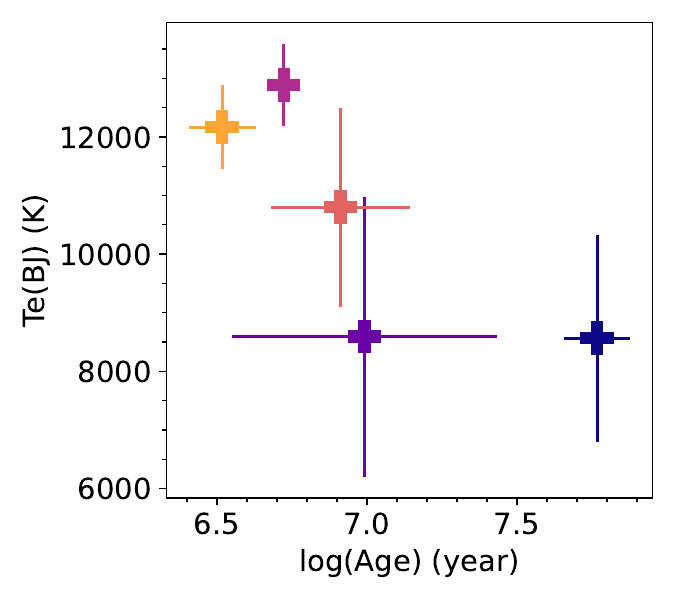}
\caption{\emph{Left:} Example of a fit to capture \etemp\ from the Balmer Jump, using the highest \ewoIII\ stack from Figure~\ref{fig:spec_wOIII}.  Data are shown in grey.  A single component stellar fit is shown in black.  A fit of stars and nebular gas emission are shown in cyan.  \emph{Right:} Comparison of the BJ temperatures and stellar population ages derived from SED modeling for all the five stacks.  The same color coding is used as in Figure~\ref{fig:spec_wOIII}. }
\label{fig:bjfit}
\end{figure*}

One of the most noteworthy features of the optical spectra of Figure~\ref{fig:spec_wOIII} is the presence of a Balmer jump, which is clearly visible in the two stacks with the highest \ewoIII, and for which we show a zoom-in in Figure~\ref{fig:bj}.  In the lower \ewoIII\ stacks (purple, blue) this break is positive, and arises because A and B type stars contribute significantly to the light at $\simeq 3650$\AA: these stars have suitable temperatures to produce a significant population of neutral hydrogen with electrons excited to the $n=2$ level.  Photoionization in the stellar atmosphere out of this state produces the stellar Balmer break.  In contrast, the higher \ewoIII\ stacks (orange, yellow) show an inverted break: this arises because of free electrons in the \hII\ regions, which can recombine into the $n=2$ or $n=3$ (or other) states, producing the Balmer and Paschen (or other) continua, respectively.  The amplitude of this break depends on the ratio of the densities of electrons with $\simeq 1$ and $\simeq 0$\:eV, and hence the gas temperature: the amplitude of the nebular Balmer jump decreases with increasing \etemp. 

Our spectra contain both stellar and nebular emission, and we therefore cannot simply measure the amplitude of the nebular Balmer jump to derive \etemp, because of the unknown contribution of the stellar break.  We therefore build a spectral model to infer \etemp, accounting for both stellar continuum, and nebular radiation from the recombination lines, free-bound, free-free, and 2-photon continua of H and He.  For starlight we take template spectra from \emph{Starburst99} \citep{Leitherer.1999,Vazquez.2005}, assuming stellar evolutionary tracks from the Geneva group, a Salpeter IMF between 0.1 and 100\,\msun\ and all available metallicities.  We fit an age parameter assuming a constant star formation rate (SFR), using the \citet{Reddy.2015,Reddy.2016} dust attenuation laws, and an arbitrary normalization parameter (as the spectra are normalized). The nebular gas is generated using the code of \citet{Schirmer.2016}, which we run on a grid spaced by 1000\,K and then interpolate using spline functions. The nebular reddening is treated across the full wavelength range, again using the \citet{Reddy.2015,Reddy.2016} law; this is the same law as used for the stellar light although stellar and nebular reddenings are allowed to be independent. The model conserves the number of ionizing photons and normalizes the spectrum to the total (maximum) photoionization, but introduces another scale factor to account for a fraction of `lost' LyC photons, which either escape the galaxy or are absorbed by dust grains. 

\begin{deluxetable*}{cccccc}[htb!]
\tablecaption{Electron temperatures and abundances estimated from the five stacks.\label{tab:abundvals}}
\tablehead{
\colhead{stack \#}  &  \colhead{\etemp (BJ, kK)}  &  \colhead{\etemp ([\oIII], kK)}  &  \colhead{\logOH}  &  \colhead{\logCO}  &  \colhead{\logNO}  }
\startdata
     $1$  &     $8.57 \pm 2.65$  &     $14.19 \pm 4.54$  &     $7.93 \pm 0.30$  &     $-1.02 \pm 0.00$  &     \nodata \\ 
     $2$  &     $8.59 \pm 3.58$  &     $15.40 \pm 1.57$  &     $7.82 \pm 0.13$  &     $-0.97 \pm 0.58$  &     \nodata \\ 
     $3$  &     $12.89 \pm 1.05$  &     $14.54 \pm 1.20$  &     $7.92 \pm 0.08$  &     $-0.57 \pm 0.32$  &     $-0.42 \pm 0.70$  \\ 
     $4$  &     $10.80 \pm 2.55$  &     $18.07 \pm 1.96$  &     $7.68 \pm 0.12$  &     $-0.94 \pm 0.49$  &     $-0.92 \pm 0.61$  \\ 
     $5$  &     $12.17 \pm 1.08$  &     $20.73 \pm 1.79$  &     $7.48 \pm 0.10$  &     $-1.00 \pm 0.22$  &     $-1.05 \pm 0.52$  \\ 
\enddata
\end{deluxetable*}

An example of this method is shown in the left panel of Figure~\ref{fig:bjfit}.  Here we show the data from the stack of the highest \ewoIII\ galaxies in gray, and the best-fitting pure stellar model is shown in black. This stellar fit well captures the very blue end of the UV ($\lambda_\mathrm{rest}<3000$\,\AA) and the red end of the optical ($\lambda_\mathrm{rest}>5500$\,\AA), but does not match the data in the vicinity of the Balmer break.  The cyan lines, however, represent a combined fit of stellar continuum and nebular gas, constrained by the Balmer lines and total continuum level: this almost perfectly reproduces the complete spectrum, including the region around the Balmer jump.  The right hand panel of the same Figure shows \etemp\ compared to the age of the stellar population from the same fit.  Temperatures are found to range from $\simeq 13$,000\,K, which are found preferentially in galaxies with ages $\simeq4$\,Myr, down to temperatures as low as 8,000\,K, which are found in galaxies with ages of $10^7$ to $10^8$\,yr.  These temperatures are reported in Table~\ref{tab:abundvals}.

\subsubsection{Forbidden Oxygen Lines}\label{sect:TeOIII}

We compute the electron temperature in the high ionization zone from the (dust-corrected) [\oIII]5008/1666\,\AA\ ratio, using version 1.1.15 of \texttt{PyNeb} \citep{Luridiana.2015}, with transition probabilities from \citet{FFT04} and collision strengths from \citet{AK99} and \citet{Kal09}.  We note here that the final \opp\ temperatures are almost twice the assume value for the intrinsic Balmer decrement, which would decrease the intrinsic \halpha/\hbeta\ ratio from 2.86 to 2.76, and require more attenuation to match the observed ratios.  However, as we show in Section~\ref{sect:bj}, the average temperatures in the nebula are more consistent with $10^4$~K when derived from radiation produced by recombining hydrogen, so we do not modify this intrinsic line Balmer decrement. 

Because of the low resolution of NIRSpec/PRISM, we do not resolve any of the density-dependent close doublets, so assume an electron density (\edens) of 250\,cm$^{-3}$ throughout for consistency with other studies at similar redshifts \citep{Sanders.2018,Reddy.2023a,Revalski.2024}. This provides the temperatures in the \opp\ zone (hereafter $T_\mathrm{high}$), which represent only a fraction of the nebula, and do not sample the likely-dominant regions where oxygen is only once-ionized.  We do not have access to the very weak auroral lines of [\oII] at 7320,7330\AA, and do not detect the [\nII]$\lambda 5575$\AA\ line to measure \etemp\ in the less ionized medium, so instead use the scaling function from \citet{Garnett.1995} to estimate the temperature of the \op\ zone (hereafter $T_\mathrm{low}$). 

\begin{figure}
\includegraphics[width=1.1\columnwidth]{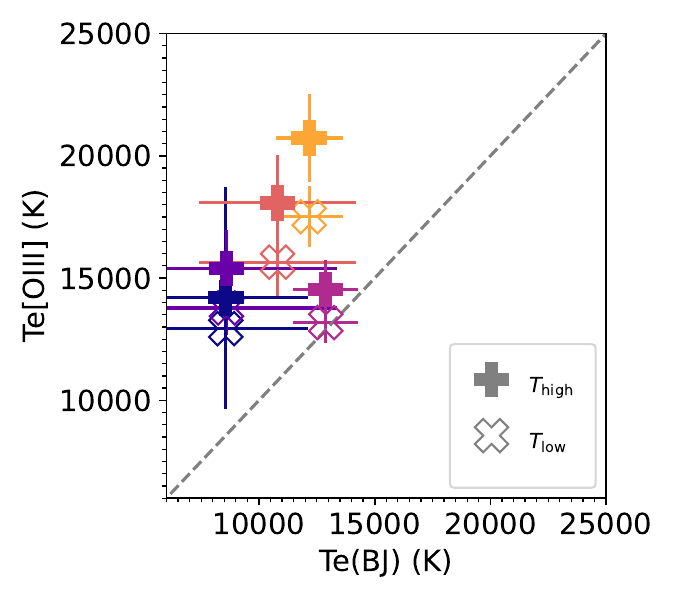}
\caption{Comparison of the [\oIII] and BJ temperatures.  Solid pluses encode the high-ionization gas measured directly from [\oIII] ratios, and open crosses show the estimated temperature in the [\oII] zone, computed using the formulae of \citet{Garnett.1995}. Color coding is the same as in Figure~\ref{fig:spec_wOIII}.  }
\label{fig:comparetemp}
\end{figure}

Following this approach we estimate the temperature of the electrons as probed by collisionally excited oxygen, which ranges between $\simeq 14,000$ and $\simeq 21,000$\,K for the \opp\ zone, and $\simeq 13,000$ and $\simeq 17,500$\,K for the \op\ zone.  We report these  temperatures in Table~\ref{tab:abundvals} and show them contrasted with those measured from the Balmer Jump in Figure~\ref{fig:comparetemp}.  There is clearly a substantial offset between the two temperatures, which amounts to $\sim 40$\% on average.  We discuss this in more detail in Section~\ref{sect:disc:temp}. 

\subsection{Metal Abundances}\label{sect:meth:abund}
The derivation of these temperatures leads us directly towards the abundances of O, C, and N, using the temperature-dependent collisional and recombination rate coefficients calculated in \texttt{PyNeb} \citep{Luridiana.2015}.  We now go through each of these elements in turn. 

\paragraph{Oxygen}
Oxygen abundances are derived using the ratio of [\oII]3727/\hbeta\ and $T_\mathrm{low}$ to obtain the abundance \op/H$^+$.  With the \op\ temperatures derived above we fix the recombination and collisional coefficients for H and O, and calculate directly the abundance that would produce the observed line ratio. By analogy, we take [\oIII]5008/\hbeta\ and $T_\mathrm{high}$ to obtain the abundance \opp/H$^+$.    We sum \op/H$^+$ + \opp/H$^+$ to obtain the total oxygen abundance O/H.  For our standard set of assumptions, the nebular oxygen abundance falls in the range $12+\log(\mathrm{O/H})=7.5$ to 8.0 (see Table~\ref{tab:abundvals}).

\begin{figure}
\includegraphics[width=0.99\columnwidth]{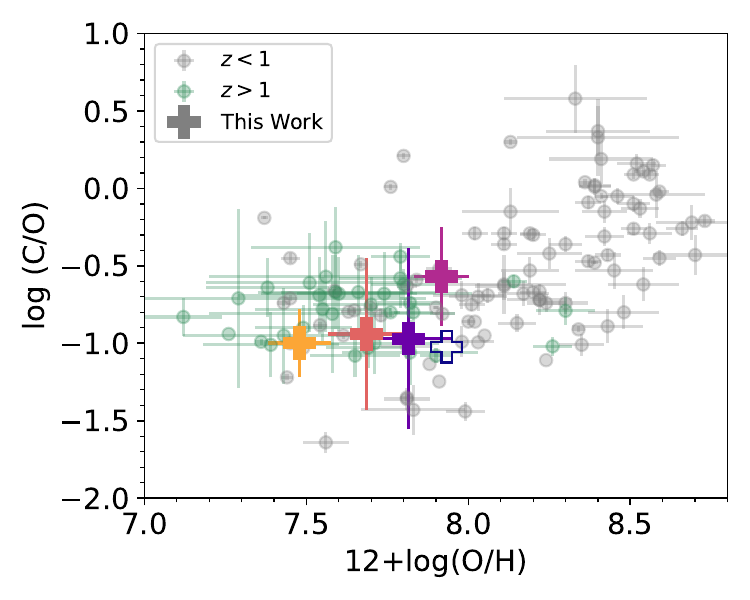}
\caption{The abundance of carbon with respect to that of oxygen (C/O) vs the nebular oxygen abundance.  Points from our five main stacks are shown in colors consistent with other figures; the open point has an uncertainty of $\simeq 1$\:dex and is not treated as a measurement.  Various literature points are also shown and labeled in the caption.  High-$z$ points are compiled from \citet{Erb.2010,Christensen.2012,James.2014a.CASSOWARY,Bayliss.2014,Stark.2014,Steidel.2016} (composite), \citet{Amorin.2017,Acharyya.2019,Mainali.2020,Arellano-Cordova.2022,Curti.2023,Iani.2023,Matthee.2021,Stiavelli.2023,Isobe.2023,Jones.2023,Citro.2024,Castellano.2024,Hsiao.2024}. Low redshift samples include \citet{ToribioSanCipriano.2016,ToribioSanCipriano.2017,Lopez-Sanchez.2007, Peimbert.2005,Esteban.2004,Esteban.2009,Esteban.2014, Garcia-Rojas.2004, Garcia-Rojas.2005, Garcia-Rojas.2007, Senchyna.2017, Pena-Guerrero.2017, Berg.2016, Berg.2019.CIVHeII,Senchyna.2021}.}
\label{fig:co_ratio}
\end{figure}

\paragraph{Carbon}
Carbon abundances are derived using the ratio of \cIII]1909/\hbeta\ and $T_\mathrm{high}$ to obtain the abundance C$^{++}$/H$^+$.  Ionization correction factors are calculated following \citet{Berg.2019.CNOdwarf} by first deriving the ionization parameter, \logU, from the O$_{32}$ ratio. For galaxies showing \cIV\ in emission we also tested the summation of the \cIV\ and \cIII] abundances to avoid the need for ICFs. We report these \logCO\ measurements in Table~\ref{tab:abundvals} and show the variation of the (C/O) ratio with \logOH\ in Figure~\ref{fig:co_ratio}.  For our standard set of assumptions, the carbon/oxygen ratio covers very little range, and all the five stacks show ratios consistent with $\log (\mathrm{C/O})\simeq -1$, regardless of O/H. 

\begin{figure*}
\includegraphics[width=9.0cm]{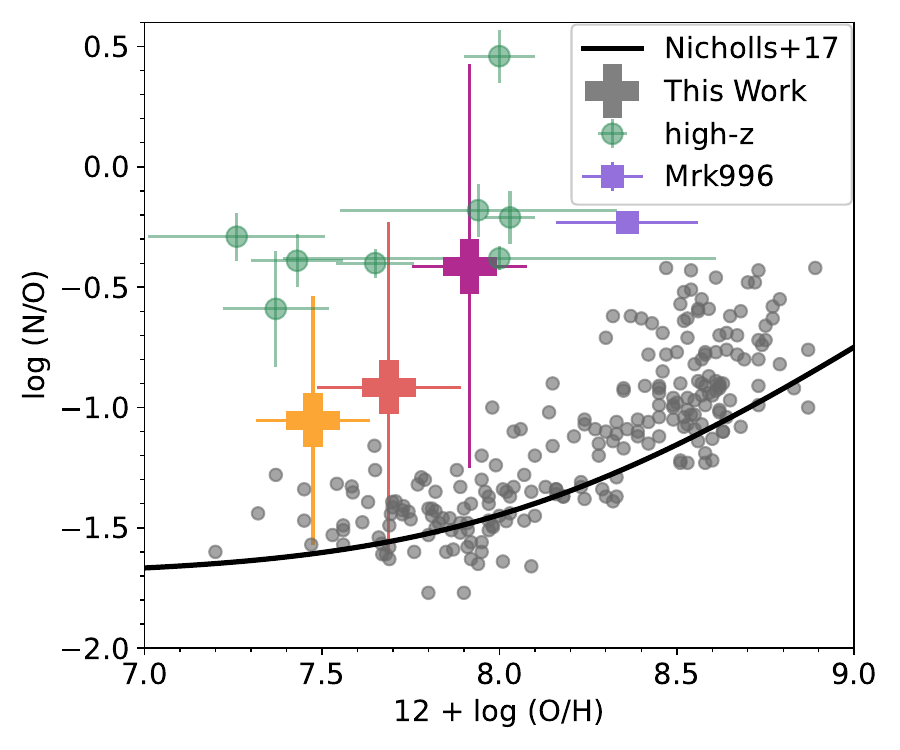}
\includegraphics[width=9.0cm]{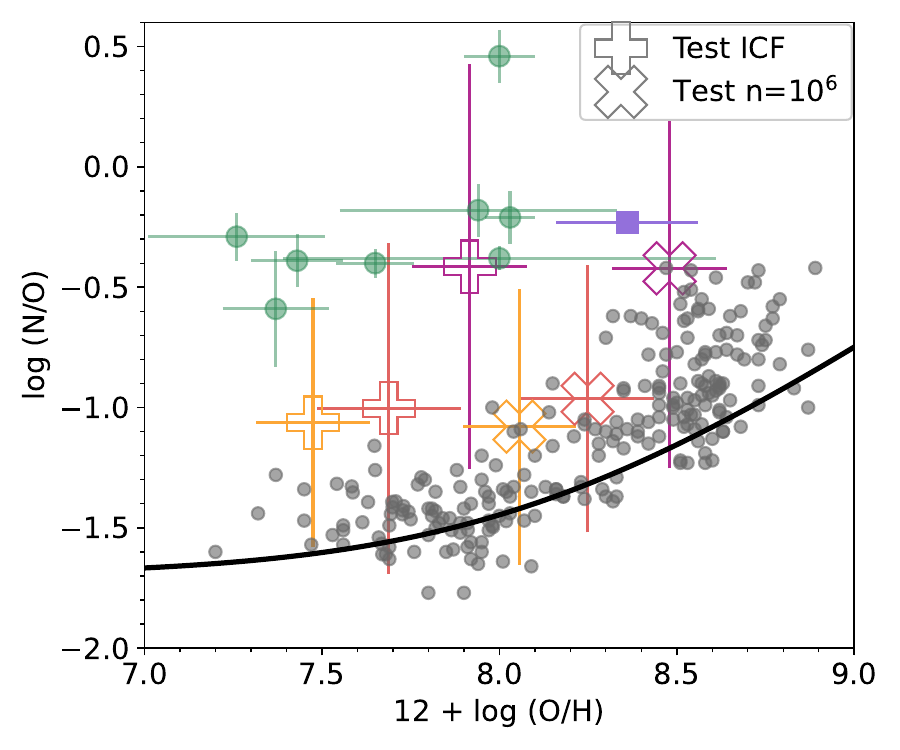}
\caption{The abundance of nitrogen with respect to oxygen (N/O) vs the nebular oxygen abundance, \logOH.  High-$z$ literature data are shown in green  \citep{Bunker.2023,Isobe.2023,Marques-Chaves.2024,Ji.2024,Castellano.2024,Napolitano.2024_GHZ9,Topping.2024,Schaerer.2024}, and the low-$z$ galaxy Mrk\,996 \citep{James.2009} is represented by the purple square.  Local galaxies, compiled from \citet{vanZee.2006} and \citet{Berg.2016,Berg.2020} are shown as gray circles, and the sample of Milky Way stars from \citet{Nicholls.2017} is shown by the black line.  The left plot shows the N/O and \logOH\ computed for our standard set of assumptions: \edens\,$=250$\,cm$^{-3}$, and without ICFs applied so that we can compare our stack measurements with literature values.  In the right plot we test the effects of the low density assumption by changing it to $10^6$\,cm$^{-3}$ (open large X-markers) and applying new ICFs (open plus markers).}
\label{fig:no_ratio}
\end{figure*}

\paragraph{Nitrogen}
Nitrogen abundances are derived using the ratio of \nIII]1749/\hbeta\ and \nIV]1483,86/\hbeta, together with $T_\mathrm{high}$ to obtain the abundance N$^{++}$/H$^+$ and N$^{3+}$/H$^+$, respectively.  We use fluxes derived from numerical integration of the emission lines.  To bypass the need for ICFs, we calculate N/O using (N$^{++}$+N$^{3+}$)/O$^{++}$ or N$^{++}$/O$^{++}$ when \nIV] was not detected.  We report these \logNO\ measurements in Table~\ref{tab:abundvals} and show the variation of the (N/O) ratio with \logOH\ in Figure~\ref{fig:no_ratio}, using our standard assumptions in the left-hand panel.  While uncertainties are large, the measurements in our stacks lie between the high-$z$ galaxies and estimates from local dwarf galaxies and extragalactic \hII\ regions.

\subsection{Tests of Fundamental Assumptions}\label{sect:testassumpt}

Without access to density-sensitive doublet like [\oII], we had to assume \edens, taking a value of 250\,cm$^{-3}$ for consistency with other high-$z$ measurements \citep[e.g.][]{Sanders.2023}. The UV lines all have critical densities in excess of $10^6$\,cm$^{-3}$, but the optical lines of [\oII]3727 and [\oIII]5008 used for both oxygen abundances and temperature measurements do not.  We therefore run all of our calculations with a high value of \edens\,$=10^6$\,cm$^{-3}$.  

This intervention has two main results.  The first is on the electron temperatures, for which $T_\mathrm{high}$ decreases systematically by $\sim 20$\%.  In fact, when calculating $T_\mathrm{low}$ using the \citet{Garnett.1995} formulation, the [\oIII] temperatures become only $\sim 20$\% hotter than those derived by the Balmer jump.  As a consequence of this, all abundances increase: \logOH\ increases by the substantial increment of 0.5\,dex, shifting all the points in Figures~\ref{fig:co_ratio} and \ref{fig:no_ratio} to the right.  Abundances of C and N, however, change by similar factors and the (C/O) and (N/O) ratios are close to constant (right panel of Figure~\ref{fig:no_ratio}).  This shifts our points in the (N/O) diagram to become more in-line with the local \hII\ regions, suggesting no enhancement of nitrogen.  Note, of course, that the individual high-$z$ galaxy estimates would shift by comparable amounts if the density were so high, implying much smaller levels of N enhancement than otherwise have been reported. 

This also brings the (C/O) ratios far from the locus of literature points, which show good continuity between the high- and low-redshift points.  In contrast to the (N/O) estimates, the local (C/O) datapoints have measured electron densities that are typically two-three orders of magnitude below the $10^6$\,cm$^{-3}$ at which these calculations have been run.  On the other hand, the low-$z$ objects typically must have more extended star formation histories on average, and we could expect (C/O) to be lower in the $z>5$ sources.

\begin{figure*}
\centering
\includegraphics[width=0.40\textwidth]{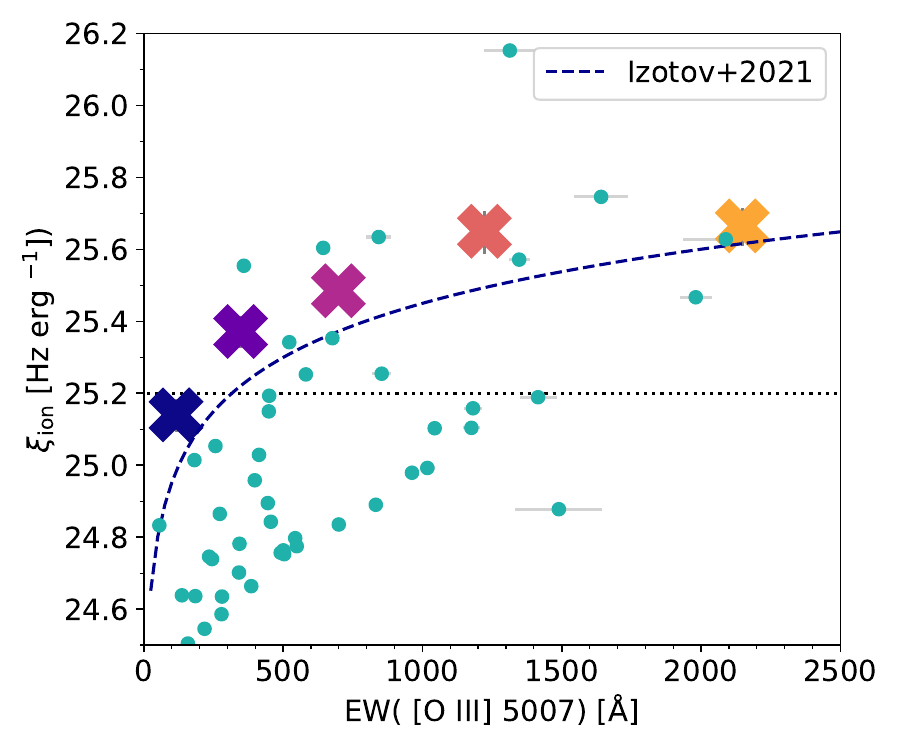}
\includegraphics[width=0.40\textwidth]{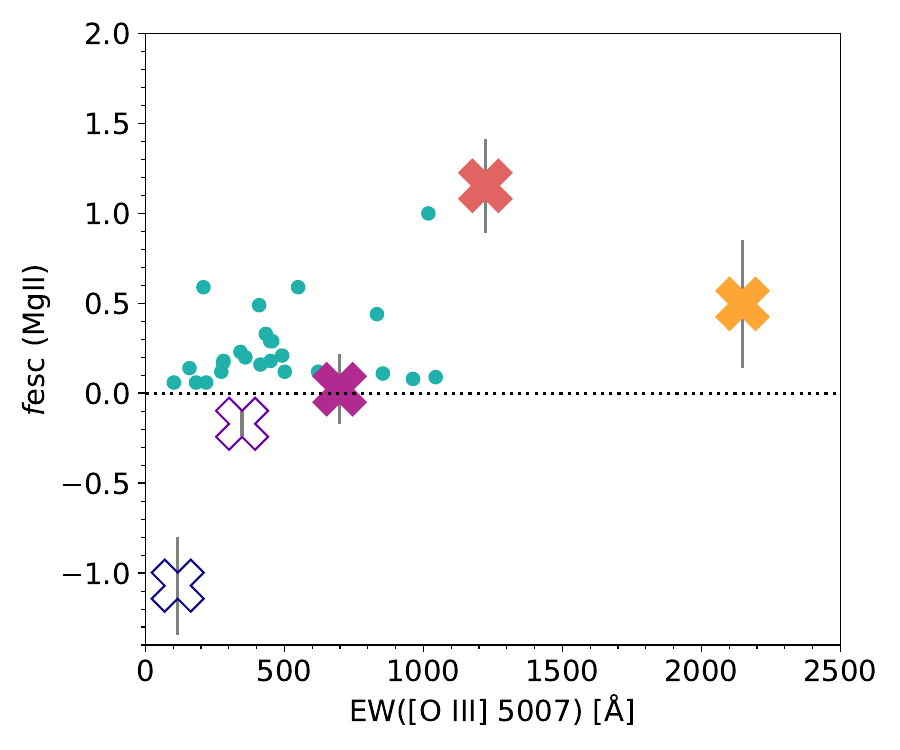}\\
\includegraphics[width=0.40\textwidth]{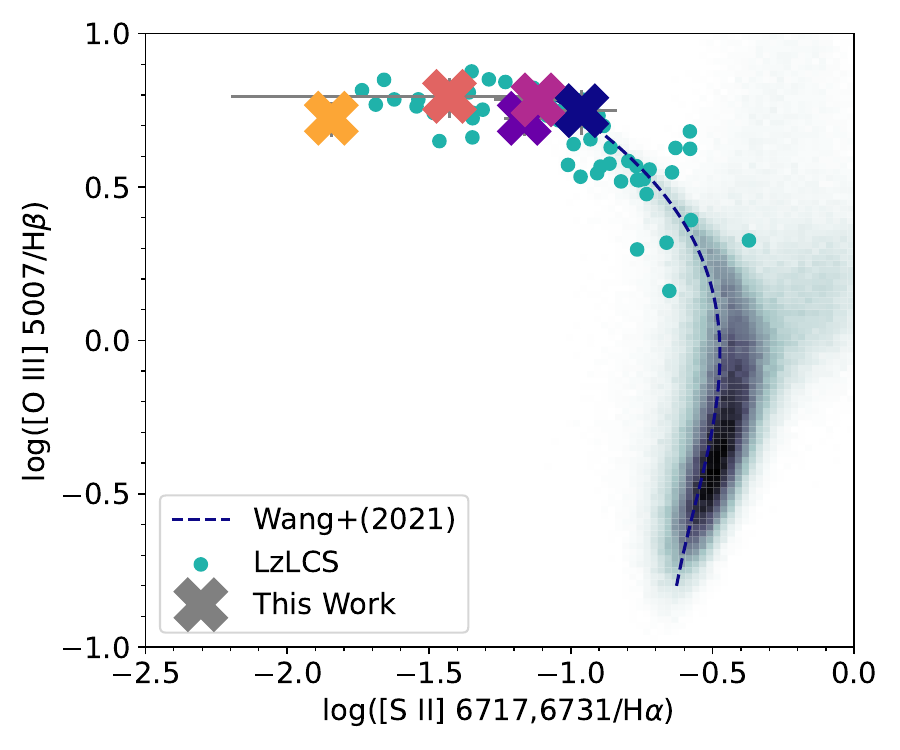}
\includegraphics[width=0.40\textwidth]{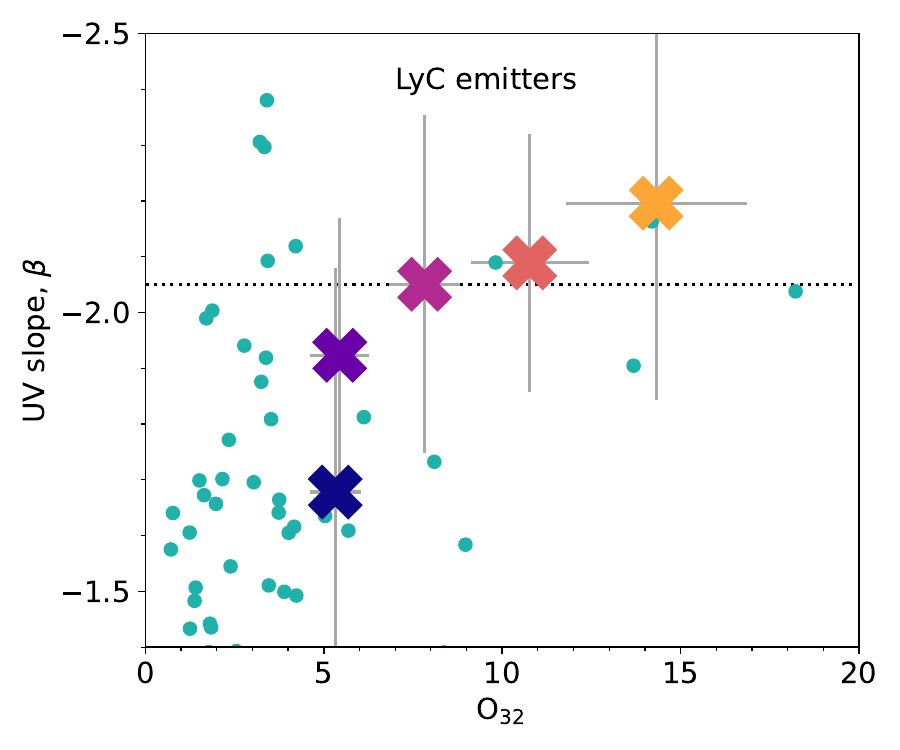}
\caption{The production and potential escape of ionizing radiation. Upper left shows the ionizing photon production efficiency, \xiion, where the canonical value needed to reionize the Universe \citep{Robertson.2015} is illustrated by the shaded region.  We project \xiion\ against \ewoIII\ for clarity, and show the empirical relation of \citet{Izotov.2021.analogs} as the dashed line.  Upper right shows the inferred escape fraction of \mgII, again against \ewoIII. Lower right shows the deficit of [\sII] emission, defined in \citet{Wang.2019,Wang.B.2021}, together with the locus of points from the Sloan Digital Sky Survey. Lower right shows the UV continuum slope $\beta$ against the O$_{32}$ ratio, with the shaded region showing the area of LyC-emitting galaxies from \citet{Chisholm.2022}.  Measurements from the stacks are presented as large crosses, following the color scheme of Figure~\ref{fig:spec_wOIII}, which we contrast with measurements from the LzLCS    shown in light green.}
\label{fig:ionrad}
\end{figure*}

\begin{deluxetable}{cccc}[htb!]
\tablecaption{Average LyC-emission quantities derived from galaxies that comprise the five stacks\label{tab:lycvals}}
\tablehead{
\colhead{stack \#}  &  \colhead{log(\xiion/[Hz erg$^{-1}$])}  &  \colhead{\fescmgii}  &  \colhead{[\sII] deficit} }
\startdata
     $1$  &     $25.14 \pm 0.04$  &     $-1.07 \pm 0.26$  &     $0.03 \pm 0.24$  \\ 
     $2$  &     $25.37 \pm 0.04$  &     $-0.17 \pm 0.06$  &     $-0.21 \pm 0.53$  \\ 
     $3$  &     $25.48 \pm 0.02$  &     $0.02 \pm 0.18$  &     $-0.06 \pm 0.30$  \\ 
     $4$  &     $25.65 \pm 0.05$  &     $1.15 \pm 0.25$  &     $-0.36 \pm 0.22$  \\ 
     $5$  &     $25.67 \pm 0.05$  &     $0.50 \pm 0.35$  &     $-0.89 \pm 0.19$  \\ 
\enddata
\end{deluxetable}

\section{Results III: Production and Escape of Ionizing Radiation}\label{sect:res:ion}
We make several inferences on the production and escape of ionizing radiation, and compute two of the main quantities in the ionizing emissivity: the ionizing photon production efficiency (\xiion), and the escape fraction of ionizing radiation (\fesclyc).  \xiion\ is defined as the number of ionizing photons produced per unit time per unit UV luminosity: $\equiv Q_0/L_\mathrm{UV}$  [Hz~erg$^{-1}$], where $Q_0$ is the number of ionizing photons produced per second for $L_\mathrm{UV}$ is the luminosity density per unit frequency \citep[e.g.][but see \citet{Tarrab.1985} and \citet{Mas-HesseKunth.1991} for some interesting historical perspectives on \xiion\ as the `$R$' and `$B$-parameter']{Robertson.2015}.  We compute $Q_0$ from the optical emission line spectrum using $Q_0 = 2.1\times 10^{12} L_\mathrm{H\beta}^0$, where $L_\mathrm{H\beta}^0$ is the dust corrected \hbeta\ luminosity in erg~s$^{-1}$. $L_\mathrm{H\beta}^0$ is computed from the \halpha\ and \hbeta\ fluxes assuming the \citet{Reddy.2015} attenuation law.  We show the resulting calculation of \xiion, projected against the \ewoIII\ variable that defines the stacks, in the upper left panel of Figure~\ref{fig:ionrad}.  log(\xiion/[Hz\,erg$^{-1}$]) ranges from 25.15 to 25.65, and increases monotonically with \ewoIII\ -- this relation has been seen numerous times in the past  \citep{Chevallard.2018, Tang.2019, Simmonds.2024a.xiionbursty}, and is expected since both EWs and \xiion\ encode line fluxes divided by stellar continuum.  Values of \xiion\ are reported in Table~\ref{tab:lycvals}.

Naturally the galaxies in our sample lie at redshifts that are far too high to measure the output of ionizing continuum, so for \fesclyc\ we must rely upon independent diagnostics.  For this we turn to the escape fraction of \mgII\ radiation \citep[\fescmgii][]{Henry.2018, Chisholm.2020, Xu.2022.lzlcs, Xu.2023.lzlcs} and the so-called [\sII]-deficit \citep{Alexandroff.2015, Wang.2019, Wang.B.2021}.  \citet[][see also \citealt{Henry.2018}]{Xu.2022.lzlcs} have published polynomial fitting functions to describe the relationship between the ratio of both components of the \mgII\ doublet ($\lambda=2796$ and 2803~\AA, independently) to the [\oIII]5008 luminosity, compared to the  O$_{32}$ ratio for both density and ionization bounded nebulae, based upon extensive sets of CLOUDY simulations.  We adopt these fits to compute the expected luminosities in the \mgII\ doublet using our [\oII] and [\oIII] measurements, and sum them to obtain the expected intrinsic \mgII.  We then obtain \fescmgii\ by dividing the measured \mgII\ luminosity by the estimated intrinsic value.   We show the result in the upper right panel of Figure~\ref{fig:ionrad}.  We make no attempt to correct for interstellar absorption of the resonant \mgII\ line which, unlike in the high-resolution ground-based observations of \citet{Xu.2023.lzlcs} we obviously cannot resolve (the effect would be to increase our \fescmgii\ estimates). We instead plot negative \fescmgii\ as open symbols.  \fescmgii\ ranges from below 0 to almost 100\%.  Values of \fescmgii\ are reported in Table~\ref{tab:lycvals}.

For the [\sII] diagnostics we follow a similar procedure, and calculate both the [\oIII]5008/\hbeta\ and [\sII]6717,31/\halpha\ ratios directly in the stacked spectra.  We then compute the [\sII]-deficit ($\Delta$[\sII]) using the formulae in \citet{Wang.B.2021}, where $\Delta$[\sII] is defined as the distance of a galaxy's measured log([\sII]/\halpha) from the expected value based upon its log([\oIII]/\hbeta) and the locus of SDSS points on the BPT diagram.  We plot these ratios in the lower left panel of Figure~\ref{fig:ionrad}, and report values of $\Delta$[\sII] in Table~\ref{tab:lycvals}. log([\oIII]/\hbeta) are relatively constant at $\simeq0.75$, but log([\sII]/\halpha) drifts monotonically away from the SDSS locus for increasing \ewoIII. Estimates of $\Delta$[\sII] are given in Table~\ref{tab:lycvals}.

We finally compute the O$_{32}$ ratio directly from the line luminosities, and UV continuum slope, $\beta$, from the stacks with luminosity normalized at 3650\,\AA\ (second left panel of Figure~\ref{fig:spec_wOIII}).  We show these in the lower right panel of Figure~\ref{fig:ionrad}.  Across the sequence of increasing \ewoIII, UV slopes become bluer and O$_{32}$ values higher.  We illustrate the LyC-leaking region of the $\beta$ range from \citet{Chisholm.2022} without shading.  $\beta$ has already been reported in Table~\ref{tab:sampdesc}.  

Throughout Figure~\ref{fig:ionrad} we contrast results from these stacks with measurements made in individual galaxies from the Low-Redshift Lyman Continuum Survey, using data from \citet{Saldana-Lopez.2022} for \xiion, \citet{Chisholm.2022} for $\beta$, \citet{Flury.2022pap1} for \ewoIII, \citet{Xu.2023.lzlcs} for \mgII, and \citet{Wang.B.2021} for [\sII].

\section{Discussion}\label{sect:disc}

\subsection{UV Line Strengths and Implications for the Ionizing Sources}\label{sect:disc:ioncond} 
Our stacks reveal an increase in the equivalent widths of rest UV lines (e.g. \cIII]) with the EW of [\oIII] in the optical (Figures~\ref{fig:spec_wOIII} and~\ref{fig:wUV_wOIII} and Section~\ref{sect:stackdescribe}).  This is a very natural result, especially given the increase in \xiion, shown in Figure~\ref{fig:ionrad} which normalizes line flux by UV continuum directly.  Similar phenomena have been seen before \citep{Berg.2016, Senchyna.2017} and our results for \cIII] agree well with those of \citet{Ravindranath.2020}, who obtained and compiled a large sample of galaxies over a range of redshifts.  This correlation was also the probable explanation for the high detection rate of \cIII] \citep{Stark.2017,Laporte.2017} in Spitzer-selected [\oIII]-emitters at $z\approx 7$ \citep{Roberts-Borsani.2016}.  

Large numbers of \cIII]-bright galaxies have also been identified at mid-$z$, most notably among the sample of 450 Main-Sequence galaxies studied by \citet{Nakajima.2018} and \citet{Llerena.2022} at $z=2-4$.  These show an average \ewcIII\ of just 2.2~\AA, which is consistent only with the least intensely star-forming subsets of our sample.  This value of \ewcIII\ is already doubled by the 2nd bin in our \ewoIII\ stacks, and exceeded by a factor of $\simeq 8$ in our most intensely star-forming stack, which shows \ewcIII~$\simeq 16$\AA.  Nakajima et al's models suggest an upper limit to \ewcIII\ of 12~\AA\ for a zero-age starburst comprised of `conventional' stars, which is exceeded here by $\sim 30$\%.  We stress again that this stacked spectrum of the highest \ewoIII\ galaxies is the average of 130 objects, and it is surprising that the \emph{average} source should lie $\sim 30$\% above this limit in \ewcIII\ if conditions in the two galaxy samples were comparable.  

Notably, the factor of $\simeq10$ increase in \ewcIII\ over our \ewoIII\ range is significantly larger than the factor of $\simeq3$ increase in \xiion\ (Figure~\ref{fig:ionrad}).  This implies that something more than a simple increase in $Q_0$ must be at play.  For example, even spectra with more evolved stellar population (showing a normal stellar Balmer break, with ages $\sim 100$\,Myr) also show \xiion\ estimates as high as 25.2.  We also see an increase in the ratio of the \cIV/\cIII] fluxes (Figure~\ref{fig:uvbpt}), which may point towards and evolution of the shape of the ionizing continuum and/or other nebular conditions (e.g. an increase in temperature associated with lower gas-phase metallicity).  Indeed, \citet{Nakajima.2018} have showed that the strength of \cIII] and \cIV\ peaks at different metallicities in their photoionization models, as a result of changes in the gas temperature and the ionization structure of the ISM.  A number of possibilities may underlie the trends seen in the stacks: (1.) lower gas metallicities may be implicated in the upward evolution of the \cIII] EW \citep[e.g.][]{Ravindranath.2020,Roberts-Borsani.2024} -- indeed we do find a general decrease in oxygen abundance across the stacks, ranging from 12+log(O/H)$\simeq 8.0$ to 7.5, which covers exactly the range where \cIII] EW should peak.  These UV lines are also extremely temperature sensitive and are thus mainly visible in low-metallicity targets: metallicity seemed the factor driving most their presence/absence in CLASSY \citep{Mingozzi.2024}. (2.) AGN may play a more dominant but currently unseen role, and their harder ionizing spectra produce more $h \nu > 24$\,eV photons than typical stars.  It is interesting in this context that the [\cII]2326 feature becomes visible only in the highest \ewoIII\ stack.  However we do not see other trends typical of AGN, and the fact that UV luminosities decrease and \xiion\ is consistent with ordinary star forming galaxies may disfavor the AGN interpretation. (3.) the higher \ewoIII\ galaxies may be the result of galaxies hosting a higher fraction of common-envelope binary stars and/or stripped stars are present in these higher \ewoIII\ stacks.  This would provide qualitatively similar changes to the ionizing spectra \citep{Gotberg.2019} as for AGN over the comparatively narrow energy range of 13.6 to 24 eV. (4.) low-velocity shocks become more prevalent \citep[e.g.][]{Jaskot.2016}, increasing the strength of semi-forbidden \cIII] compared to the continuum.  
    
Interestingly, the behavior of the \cIV\ emission is not nearly as extreme: it peaks at only \ewcIV\,$\simeq 3$\,\AA.  This is $\lesssim 1/2$ the strength of the \cIV-emitter sample of VANDELS galaxies presented in \citep{Mascia.2023}, which have a median \ewcIV\ of 7.6\,\AA.  We note, of course, that statistics taken on a \cIV-emission only sample will be biased high by excluding the dominant fraction of galaxies that do not have \cIV\ fluxes above the detection threshold.  We finally remake that, while the \cIII] EWs of our stacks are extreme by the standards of mid-$z$ individual galaxies, the \cIV\ emission appears not to be.  

\subsection{Gas Temperatures in Early Galaxies} \label{sect:disc:temp}
Electron temperatures derived from the ratio of [\oIII]5008/1666\AA\ emission lines fall in the range 14,000 to 21,000\,Kelvin and increase monotonically with the intensity of star formation within galaxies: this change occurs as ages decrease from $\simeq 60$\,Myr to just 3\,Myr (right panel of Figure~\ref{fig:bjfit}).  This trend in temperature is expected as episodes of star formation become younger and contain a higher fraction of more massive stars (although see the caveats concerning AGN contamination above): ionization by stars that are hotter on average transfer more kinetic energy to the free electrons in the photoionization processes.  At the same time, cooling is less efficient at lower metallicities, and we also see a clear change in the nebular oxygen abundance, also decreasing as temperature increase (and ages decrease), as shown in Figure~\ref{fig:co_ratio}. 

Given that we expect there to be a distribution of temperatures in the \hII\ regions, we would further predict that \etemp(BJ)  
should be lower than other estimates, as seen extensively in planetary nebulae and \hII-regions \citep[e.g.][]{Wesson.2005}.  Using only the [\oIII] line ratios to derive \etemp\ from collisional tracers, we can only measure \etemp\ for the \opp\ regions and have no information on the \op\ that dominates the total recombination rate of the nebula  (the volume of the \opp\ zone is $\sim 1/10$ that of the \op\ zone for a Str\"omgren sphere surrounding an O star, and smaller for later type stars).  We therefore scale down our estimates of \etemp\ to estimate that of the \op\ zone, which should be almost identical in size to the \hII\ region, using the prescription of \citet{Garnett.1995}.   We determine that \etemp\ derived from collisional lines is $\sim 40$~\% hotter than that derived from hydrogen recombinations, even when the correction to the low ionization zone is applied.  We question, therefore, whether one of our temperature diagnostics may be inaccurate.  

\textit{Are the Balmer jump temperatures too low?}  Regarding the method, it is hard to see how this could be the case.  The normalization of Paschen continuum is very tightly constrained by Balmer line fluxes over exactly the same wavelength range: the temperature dependence of both bound-bound and bound-free processes are treated consistently, and dust is applied to the both nebular components together.  It is hard, therefore, to get the level of the Paschen continuum wrong by significant factors.  This means the relative level of Balmer continuum on the blue side of the break, which is very clearly seen by eye (it even affects the UV slope and stellar age inference), can be tightly constrained, especially for the highest two \ewoIII\ stacks \citep[see, e.g.][]{Langeroodi.2024, Roberts-Borsani.2024}.  If we recovered an evolved stellar population, with strong stellar jump, this would increase the amplitude of the nebular break needed to match the data, and lower the temperature.  However, the recovered stellar age is just 3\,Myr for the youngest stack, and contains no significant stellar Balmer Jump (Figure~\ref{fig:bjfit}), and any deviation from this would drive temperature down and further exacerbate the discrepancy.  

\emph{Are the oxygen temperatures too high?}  One possibility is that, when using the [\oIII]5008/1666\,\AA\ ratio, small uncertainties in the dust attenuation lead to large over-corrections and \opp\ temperatures that are too high.  However, we believe this not to be the case for at least three reasons. Firstly, we can also measure \etemp\ in three of the stacks using the [\oIII]5008/4363\,\AA\ ratio: this is not nearly as sensitive to attenuation law variations, and yields temperatures consistent with those from [\oIII]5008/1666\AA.  We also have \heII\ measurements at both 1640\AA\ (Balmer alpha) and 4687\AA\ (Paschen alpha), which are very close in wavelength to the 1666 and 5008\AA\ [\oIII] lines.  If we instead take the ratio of [\oIII]/\heII\ ratios (\oIII]1666/\heII1640)/([\oIII]5008/\heII 4687) any effect of dust attenuation is canceled.  Thus we believe over-correction for dust is not responsible for the high temperatures.  

We have also assumed a fixed electron density of \edens\,$=250$\,cm$^{-3}$ throughout these calculations, and in principle the [\oIII]5008/1666\,\AA\ ratios could be artificially high if the 5008\AA\ line is collisionally de-excited.  However, the critical density of the \opp\ $^1$D$_2$ level is $6.4\times 10^{5}$\,cm$^{-3}$, which is far higher than expected for the ISM density of the average galaxy.  

Our estimates by both methods also broadly agree with other measurements made at similar redshifts and in analog galaxies.  \citet{Schaerer.2022b.1stJWST} find \etemp\,$=16,000 - 18,000$\,K from [\oIII]4363 for the two $z=7.7$ galaxies, \citet{Trump.2023} obtain \etemp\,$\simeq 20$,000~K from [\oIII] at $z=8.5$, and \citet{Curti.2023} find some temperatures in excess of 25,000\,K.  \citet{Nakajima.2023} carefully re-analyze several previously published samples of high-$z$, NIRSpec-derived \etemp\ measurements, finding many have been over-estimated, but still report temperatures as high as $\simeq 21$,000\,K for a number of systems. Similar numbers and ranges are reported by \citet{Sanders.2024}, which are all consistent with our [\oIII] results in samples of what should be the most highly ionized, hottest starbursts.  In the low-$z$ analog starburst galaxies, \citet{Mingozzi.2022} find temperatures in the range 12,000 to 20,000\,K using the UV lines for the CLASSY sample, while temperatures in excess of 19,000\,K are also reported for LzLCS using the optical lines \citep{Flury.2022pap1}.  Nothing about our [\oIII] temperature is surprising. 

Few examples of Balmer jump temperatures exist at comparable redshifts.  \citet{Welch.2024} measured Balmer Jump temperatures of 12,000\,K in the dominant LyC-emitting cluster of the Sunburst arc \citep{Rivera-Thorsen.2019}, which is consistent with our BJ estimates for the strongest [\oIII]-emitters.  We note that they also measured [\oIII] optical temperatures that were not as extreme as we found (i.e. the discrepancy is smaller) but we also point out that this cluster is thought to be extreme, showing strong nitrogen enrichment, very high densities, and significant UV emission lines \citep[][see below]{Mestric.2023,Pascale.2023}.  On the other hand our Balmer Jump temperatures are in stark disagreement with those recently reported by \citet{Katz.2024}, who find temperatures of 20,000\,K and sometimes much higher when using a similar approach, and whose sample must contain some of the same objects as ours.  It is plausible that our stacking of hundreds of galaxies picks out a more representative and less extreme sub-sample, with lower average \etemp\ than those of \citet{Katz.2024}.  While not derived from the Balmer Jump, very interesting estimate of the \op\ zone temperature has was reported by \citet{Sanders.2023}. They measured \etemp\ of 12,000 and 9000\,K using using auroral lines of [\oII] in 2 galaxies at $z=2$, where the line formation should occur in the same zone as our recombination continuum based estimates.  

Temperature fluctuations are invoked to explain abundance discrepancies in planetary nebulae and \hII\ regions (comparing collisional metal lines with recombination metal lines, in the very local Universe only; E.g. \citealt{Mendez-Delgado.2023}). If the gas is clumpy and has uneven illumination, then the distribution of temperatures will be set as the cooling rate varies.  Only the hotter parts of the nebula contribute to the production of auroral line radiation, skewing the temperature upwards (and the inferred oxygen abundance downwards).  One may expect similar behavior in the integrated spectra of galaxies, but this was not the result found by \citet{Guseva.2006}.  These authors looked at local blue compact galaxies, finding good agreement between [\oIII]5008/4363 and Balmer (and Paschen) Jump temperature estimates.  It is plausible, of course, that some of the \citet{Guseva.2006} result could be explained by aperture effects: if the brightest optical point sources are observed with narrow slits the spectra are likely biased towards the hottest regions of the nebulae, revealing high temperature that are also matched by the Balmer Jump.  However, the larger, cooler, lower ionized zones that still contribute to the free bound continuum would not be captured by small slits and large discrepancies could be avoided.  Our spectroscopic apertures, which have sizes of several kpc, would capture the full extent of the nebulae where temperatures are genuinely lower than in the small sub-regions that dominate the [\oIII] emission.

\subsection{Metal Abundance Patterns}\label{sect:disc:abunda}

\subsubsection{Carbon}
Four of the stacked spectra show log(C/O)\,$\simeq-1$, with uncertainties as low as $\simeq0.2$~dex for the highest \ewoIII\ (youngest) starbursts, where the S/N ratio of the UV lines is highest.  This uncertainty increases for the lower values of \ewoIII\ where the lines become weaker.  The central bin in \ewoIII\ scatters upwards to log(C/O)\,$=-0.6\pm 0.3$, but is not significantly different when accounting for the uncertainty. These measurement are broadly consistent with those in the literature for low metallicity galaxies at low-redshift, and also some observed at $z=6-9$ using recent JWST spectroscopy (Figure~\ref{fig:co_ratio}). Our C/O abundances also agree with the average value by \citet{Hu.2024} from the stacked analysis of 63 JWST galaxies at $5.6 < z < 9$, using higher resolution data.

The C/O ratio should increase with time during the evolution of a galaxy: it starts low (log(C/O)\,$\sim-1.5$), where the ratio simply reflects the yields of core-collapse supernova (CCSN).  CCSN explode almost immediately after an episode of star formation, and produce the overwhelming majority of oxygen in our universe, with yields that are reasonably certain.  However, the bulk of carbon is produced by low and intermediate mass stars, as they ascend the Asymptotic Giant Branch (AGB) on timescales of 100\,Myr to several Gyr.  Thus at time goes on, the (C/O) ratio increases from some plateau value as proportionally more carbon than oxygen is delivered to the ISM. 

Our measurements of the oxygen abundance (Figure~\ref{fig:co_ratio}) absolutely show the expected increase in \logOH, which grows from $\simeq 7.5$ at age of 3\,Myr to $\simeq 8$ at age of 60\,Myr (See Figure~\ref{fig:bjfit} for ages). Interesting, however, our (C/O) measurements \emph{do not capture the expected buildup of carbon}: log(C/O) appears fixed at $-1$ for all metallicities (ages) where it can be measured.  A large number of literature points from both low- and high-redshifts are over-plotted to illustrate the known variations in C/O from other samples.  This is likely because the light-weighted age we recover (e.g. Figure~\ref{fig:bjfit} extend only to ages of $\simeq10$\:Myr for stacks in which C/O can be measured -- these ages are far from sufficient to capture the major fraction of carbon production, which is highest some 500\:Myr after major episodes of star formation \citep{Mattsson.2010}.  Of course star formation history adds a more nuanced view than the simple quantity of `age', and we have made no efforts to reconstruct more detailed SFHs in this paper, nevertheless, if a significant fraction of the stellar population has formed in the last half billion years, which seems plausible for strong line-emitting galaxies at $z\simeq 5.5$ on average, then then it is hard to avoid the under-production of carbon with respect to oxygen. 

Our uncertainties on the carbon abundance are relatively small ($\simeq 0.3$\,dex) for the highest \ewoIII\ stack, and the the points do not significantly scatter -- it is likely our data have sufficient S/N to detect variations if they are present.  In fact, the \cIII] and \cIV\ lines used for the carbon abundances are of \emph{higher} S/N than the UV \oIII] lines used to estimate temperatures.  We note once more that over-corrections for dust attenuation are unlikely to mask variations in C/O for the same reason they did not affect the temperatures: dust attenuation is small, and results do not change when \heII\ is used to remove the effects of an uncertain UV-optical reddening law.  We also believe the ionization correction factor is not the culprit for this lack of evolution.  We have computed ICFs from the O$_{32}$ ratio following \citet{Berg.2019.CNOdwarf}, finding that this makes little difference.  In fact, for the highest metallicity stacks where we suspect (C/O) could be higher, the absence of an ICF would make less difference as it is unlikely/impossible for significant amounts of carbon to reside in more highly ionized states.

\subsubsection{Nitrogen}
We show the N/O ratios in Figure~\ref{fig:no_ratio}.  Recent papers by \citet{Topping.2024} and \citet{Schaerer.2024} show compilations of N/O measurements vs. 12+log(O/H) for a sample of local galaxies including CLASSY and the starburst observed by \citep{Senchyna.2022}, as well as a handful of individually detected `nitrogen-loud' galaxies at high-$z$.  We show these systems as literature points in Figure~\ref{fig:no_ratio}, together with the measurements in our stacks.  In the left panel of the figure, we provide measurements using basic ionization correction factors, for the sake of comparison with the literature points.  Our stacks cover the same range in oxygen abundance as the individual sources.  Our estimates of log(N/O) fall in the range $\simeq -0.5$ to 0.0 in the stacks for which \nIII] and/or \nIV] can be detected.  This is in excellent agreement with the measurements made at high-$z$, as well as local starburst Mrk\,996 which have broadly similar oxygen abundances.  Within the uncertainties, we do not find evolution of N/O with 12+log(O/H) although the estimate values do correlate.  Since our estimates use similar methods to those of the $z>6$ galaxies, we must conclude that if these individually detected objects show `exotic conditions', then the same exotic conditions must be found in the average galaxy over the redshift range of this survey.  

Regarding the cause of the strong nitrogen enrichment in galaxies, we can now reject explanations in order of increasing implausibility. This means we can safely rule out the possibility of the enhanced nitrogen emission being associated with tidal disruption events, since  it is implausible that the median galaxy hosts one ongoing TDE.  For similar reasons, it would be a bold claim that AGN are the culprits: while they have been recently discovered in very high numbers at these high redshifts \citep{Maiolino.2023agn,Harikane.2023agn,Hayes.2024,Kocevski.2024,Matthee.2024} there is an enormous difference between an AGN occupation fraction of a few percent and the $>50$\% needed to be visible in a median stack.  Moreover, stronger emission line diagnostics in the UV and optical do not show substantial evidence for AGN in these stacked spectra, although we note again that the [\cII]2326 line is only visible in the stack with the highest N/O.

For scenarios we cannot reject, the formation of extremely massive stars in dense stellar clusters \citep[e.g.][]{Charbonnel.2023} is an intriguing possibility, since it is plausible that selection of galaxies by strong [\oIII] emission at $z=4-10$ must bias the sample towards the most extremely star forming galaxies, with very denser interstellar media on average.  These would likely be the formation sites of the most massive stellar clusters, which are needed for stellar mergers and also nitrogen retention.  \citet{Nandal.2024} have recently calculated the abundances of stellar ejecta for fast-rotating low-metallicity stars sampled from a top-heavy IMF, finding that nitrogen abundances agree closely with recent observations of nitrogen-loud galaxies.  Their highest N/O ratios of $\simeq -0.35$ occur at \logOH\:$\simeq7.8$, which is similar to the values reported in our stacked spectra.  Of course one would not expect the entire ISM of every galaxy to be be N-enhanced to the same degree, and some mixing of the products of fast rotators and more ordinary stars would be expected, and would lower N/O at fixed \logOH, consistently with the observations. 

We also compute nitrogen abundances using the updated ICFs of Martinez et al (2024, in preparation), which are built upon the models of \citet{Berg.2019.CNOdwarf}.  These ICFs are based upon ionization parameters derived from the O$_{32}$ ratio, but over the range of ionization parameters assessed there is negligible difference between the results in Figure~\ref{fig:no_ratio}. We also test the assumption of very high gas densities, by increasing \edens\ to $10^6$ throughout the calculations, and show the result in the right-hand panel of Figure~\ref{fig:no_ratio}.  This has the curious effect of shifting our stacked measurements to be consistent with the local Universe points.  However the overwhelming majority of this effect comes from a shift in \logOH, which increases by about 0.5\:dex as the optical lines of both [\oII] and [\oIII] are collisionally deexcited.  This would also shift the literature (purple) points towards higher O/H by the same 0.5\:dex, and would seriously revise most high-$z$ metallicity estimates.

\subsection{Production and Emission of Ionizing Radiation}\label{sect:disc:ionrad}
As shown in Figure~\ref{fig:ionrad}, \xiion\ increases by a factor of $\simeq3$ across our stacks, from log(\xiion/[Hz erg$^{-1}]=25.15$ to 25.65.  These estimates are perfectly in line with those reported by \citet{Stefanon.2022} and \citet{Simmonds.2024a.xiionbursty} for emission line galaxies at similar redshifts (and indeed include some of the same galaxies used by \citealt{Simmonds.2024a.xiionbursty}).  Our estimates of \xiion\ slightly exceed those from \citet{Simmonds.2024c.xiionall} for the broader sample of JADES galaxies, and likely for reasons outlined therein: our demand for measured emission lines (for accurate spectroscopic redshifts) biases the sample towards higher values of \xiion.   The higher end of our range is consistent with the value of 25.8 reported by \citet{Atek.2024} for strongly gravitationally lensed galaxies: our least luminous stack (highest \ewoIII) has $M_\mathrm{UV}=-18.5$, which is still a factor of $\sim 10$ more luminous than those in the Atek et al sample and consistent with the expectation that \xiion\ increases further towards lower masses and luminosities (although see also \citealt{Simmonds.2024c.xiionall}).  Even if we cannot provide evidence for the star formation in the high-\ewoIII\ galaxies being bursty, the stellar populations of these galaxies are consistent with ages of just a few Myr (Figure~\ref{fig:bjfit}) and these values of \xiion\ are sufficient to reionize the Universe if the  escape fractions are high enough. 

The remaining three panels of Figure~\ref{fig:ionrad} all show that the escape fractions of ionizing radiation are probably high.  Firstly, even the plot of \xiion\ (left) shows the high-$z$ sources produce more ionizing photons per UV luminosity than their low-$z$ counterparts: we compare the JWST measurements with those from the \emph{Low Redshift Lyman Continuum Survey} \citep[LzLCS][]{Flury.2022pap1,Flury.2022pap2}.  Recent modeling results in LzLCS show that the highest \fesclyc\ values are reported for the youngest galaxies, implicating the radiative feedback that occurs only at very young ages \citep{Carr.2024, Flury.2024}. Radiative feedback in these high-$z$ galaxies is likely to be higher still, further elevating LyC emission. 

We consider four independent diagnostics of LyC emission, which have been extensively studied.  Panels 2,3 and 4 of Figure~\ref{fig:ionrad} all show that \fesclyc\ is likely to be high in the strongest star-forming two stacks.   Firstly, these stacks show escape fractions of \mgII\ in the range 50--100\%.  \fescmgii\ has been verified to closely correlate with the escape fraction of \lya\ \citep{Henry.2018,Xu.2023.lzlcs}, as its resonance nature implies the emission process of both lines will be subject to similar radiative transport effects.  A direct correlation between \fescmgii\ and \fesclyc\ has been more illusive, as also concluded by \citet{Izotov.2022}, so we cannot use the \mgII\ escape as a direct predictor or LyC output. However, we note our highest two \ewoIII\ stacks show much large median \fescmgii\ than the average galaxy in LzLCS. 

We also show the [\sII]/\halpha\ vs. [\oIII]/\hbeta\ BPT diagram in the lower left panel of the same Figure. This diagnostic is motivated by \citet{Alexandroff.2015} and \citet[][see also \citealt{Pellegrini.2012}]{Wang.2019}: the ionization potential range of singly ionized sulfur is 10.4--23.3~eV, causing the [\sII] emitting regions to lie towards the outer edge of the Str\"omgren sphere.  Truncated \hII\ regions, therefore, will show a deficit of [\sII] emission compared to line ratios that trace more highly ionized gas (e.g. the [\oIII]/\hbeta\ ratio on the x-axis).  The locus of SDSS galaxies is illustrated in this plot, together with the best fit to these objects from \citet{Wang.B.2021}.  The first three of our high-z stacks lie close to the SDSS ridge line and \citet{Wang.B.2021} prescription, albeit at the very high end in the [\oIII]/\hbeta\ plane.  They occupy a similar range in the parameter space to the more extreme half of galaxies in LzLCS in [\oIII]/\hbeta\ space.  However two the highest \ewoIII\ stacks fall far from the Wang et al. line, by 0.36 and 0.89 dex, where \emph{the average galaxy}  exceeds the most extreme deficit seen in the LzLCS program ($\Delta$[\sII] = 0.66).  As with the \mgII\ discussion above, correlations between $\Delta$[\sII] and \fesclyc\ exhibit too much spread to use for quantitative estimates, but we note that if we take the LzLCS objects for which $\Delta$[\sII] exceeds 0.3~dex then the mean \fesclyc\ reaches a significant value of 12\%.  

We finally show the variation of the UV slope ($\beta$) and O$_{32}$ ratio in the lower right panel of Figure~\ref{fig:ionrad}.  There is a monotonic bluening of the continuum color and increase in the ionization parameter across the sequence of \ewoIII.  Again, these tell the story of stellar populations becoming younger (verified by the spectral fits in Figure~\ref{fig:bjfit}), less dust obscured (verified by the \halpha/\hbeta\ ratios), and lower metallicity (not verified by stellar metallicities, but demonstrated in the gas phase by \logOH; Figure~\ref{fig:co_ratio}).  The $\beta$ values measured in the highest \ewoIII\ stacks fall into the range where \citet{Chisholm.2022} estimate \fesclyc\ to be in the 5--10\% range.  These two stacked spectra show O$_{32}$ ratios between 10 and 15, at which range \citet{Flury.2022pap2} would have found a LyC detection rate above 50\% for low-$z$ analog galaxies.  Combined with the high ionizing photon production efficiencies discussed above, we expect these samples to be very bright in the Lyman continuum.

\section{Summary and Conclusions}\label{sect:conc}

We have studied $\simeq 1000$ ostensibly star-forming galaxies at redshifts between 4 and 10 that were observed with JWST/NIRSpec in the low-resolution PRISM mode.  We performed stacking analyses on the spectra, dividing galaxies into five sub-samples sorted by their equivalent width in the [\oIII]5008 line. These combined spectra  reach total integration times of $\simeq 2-5$ million seconds per spectral pixel.  This provides the depth needed to reveal faint ultraviolet emission lines in  [\cII], \cIII], \cIV, \heII, \oIII], \nIII], \nIV] and \mgII\ which are not regularly observed in JWST spectra of high-$z$ galaxies.  We also detect, at very high signal-to-noise, a complete suite of optical emission lines at wavelengths between [\oII] and [\sII].  

We measure the luminosity and restframe equivalent width of these emission lines, paying particular attention to the UV.  We show that all features in the UV increase in strength as a function of [\oIII] EW, and much of this evolution can naturally be attributed to the decreasing average age of the stellar population with increasing \ewoIII.  We place the measurements from our five individual stacks on UV diagnostic diagrams that show line ratios and EWs, concluding they are predominantly star-forming, although we do find evidence that the spectral slope of the extreme ultraviolet becomes steeper towards younger galaxies. 

We measure the interstellar gas temperatures using the [\oIII]5008/1666 line ratios, deriving temperatures in the range 17,000 to 22,000~Kelvin.  However our very deep stacked spectra also well-expose the nebular Balmer jump in the optical, which is clearly seen and allows an independent measurement of the average electron temperature.  The \etemp\ is $\sim 40$\% lower than that derived from the [\oIII] ratio, a difference that is larger than would be expected from typical temperature differences from the \opp\ and \op\ zones. 

We use these temperatures to measure accurate abundances for oxygen, carbon, and nitrogen, finding oxygen abundances in the range $12+\log(\mathrm{O/H})=7.5 - 8.0$.  Consistent with other studies at the same redshift, we find carbon/oxygen ratios at log(C/O)$=-1$; however less consistently, we find no evidence of an increase in C/O with O/H, despite the stellar populations of the more metallic galaxies also being shown to be more evolved.  This may plausibly be explained by us catching galaxies in a strong starburst phase (demanding [\oIII] detections for redshifts), but does not erase the question of why previous studies do not find the same lack of evolution.   

Nitrogen appears to be somewhat enriched, and we find values of log(N/O)$\simeq-0.3$ in the three highest \ewoIII\ stacks, where \nIII] and \nIV] are detected.  This is consistent with recent estimates in individual galaxies.  The individual galaxies have been discussed extensively in many recent papers, and attributed to hidden AGN, tidal disruption event, or the very massive stars that may signpost the formation of proto-globular clusters.  Regardless of the exact origin of these emission lines, we believe the same processes are running at a lower level in the average galaxy at these redshifts.  We expect this is due to stellar activity, but cannot entirely rule out that it is relate to the presence of SMBHs. 

We investigate various signatures of the production and escape of Lyman continuum emission, showing that the ionizing photon production efficiency varies between log(\xiion/[Hz\:erg$^{-1}$])\,$=25.15$ and 25.65.  Concerning escape fractions we focus upon the \mgII\ and [\sII] emission lines, [\oIII]/[\oII] ratios and UV continuum color.  We compute the \mgII\ escape fraction, and show it varies between zero (and $<0$, as \mgII\ is seen in absorption in the weakest emission stacks) and almost 1, with \fescmgii$\sim 0.5$ for the two strongest stacks.  In the same two stacks the [\sII]/\halpha\ ratio falls rapidly away from the locus of local galaxies, O$_{32}$ ratios increase to above 10, and UV slopes become bluer than $\beta=-2$. All these diagnostics have been connected with high \fesclyc\ in the local Universe, and we confidently believe the highest \ewoIII\ galaxies in our sample are strong LyC-leaking galaxies. 


\begin{acknowledgements}
M.J.H. is supported by the Swedish Research Council (Vetenskapsr\aa{}det) and is Fellow of the Knut \& Alice Wallenberg Foundation. B.L.J. is thankful for support from the European Space Agency (ESA).  We thank the anonymous referee for careful reflections that have improved the quality of the manuscript, C. Simmonds for useful discussions on the ionizing photon production efficiency, and E.C. Herenz for useful feedback concerning Balmer Jump temperatures.
\end{acknowledgements}


\end{document}